\documentclass[aps, pra, showpacs, twocolumn, amsfonts, amsmath, amssymb, superscriptaddress, floatfix]{revtex4-1}

\usepackage[T1]{fontenc}
\usepackage[latin9]{inputenc}
\usepackage{color}
\usepackage{graphicx}
\usepackage{ulem}

\def\be{\begin{equation}}
\def\ee{\end{equation}}
\def\bea{\begin{eqnarray}}
\def\eea{\end{eqnarray}}

\newcommand{\vect}[1]{\mathbf{#1}}

\begin{document}

\title{Anderson localization of a Rydberg electron along a classical orbit}

\author{Krzysztof Giergiel} 
\affiliation{
Instytut Fizyki imienia Mariana Smoluchowskiego, 
Uniwersytet Jagiello\'nski, ulica Profesora Stanis\l{}awa \L{}ojasiewicza 11, PL-30-348 Krak\'ow, Poland}

\author{Krzysztof Sacha} 
\affiliation{
Instytut Fizyki imienia Mariana Smoluchowskiego, 
Uniwersytet Jagiello\'nski, ulica Profesora Stanis\l{}awa \L{}ojasiewicza 11, PL-30-348 Krak\'ow, Poland}
\affiliation{Mark Kac Complex Systems Research Center, Uniwersytet Jagiello\'nski, ulica Profesora Stanis\l{}awa \L{}ojasiewicza 11, PL-30-348 Krak\'ow, Poland
}


\begin{abstract}
Anderson localization is related to exponential localization of a particle in the configuration space in the presence of a disorder potential. Anderson localization can be also observed in the momentum space and corresponds to quantum suppression of classical diffusion in systems that are classically chaotic. Another kind of Anderson localization has been recently proposed, i.e. localization in the time domain due to the presence of {\it disorder} in time. That is, the probability density for the detection of a system at a fixed position in the configuration space is localized exponentially around a certain moment of time if a system is driven by a force that fluctuates in time. We show that an electron in a Rydberg atom, perturbed by a fluctuating microwave field, Anderson localizes along a classical periodic orbit. In other words the probability density for the detection of an electron at a fixed position on an orbit is exponentially localized around a certain time moment. This phenomenon can be experimentally observed.
\end{abstract}

\maketitle

\section{Introduction}

Transport of a particle in the presence of a disorder potential can stop totally due destructive interference effects. This phenomenon, which is accompanied by exponential localization of eigenstates of a system in the configuration space, is the famous Anderson localization \cite{Anderson1958}. It can occur in a variety of different disordered systems ranging from acoustic waves to matter waves of ultra-cold atomic gases \cite{tiggelen99,Lagendijk2009,MuellerDelande:Houches:2009}. 

Classical particle can perform diffusive motion in the phase space if dynamics of a system is chaotic. In the quantum description one can observe Anderson localization in the momentum space that is induced not by external disorder but by underlying chaotic classical dynamics. Such a dynamical Anderson localization, predicted in the kicked rotor system, is extensively investigated both theoretically and experimentally \cite{Moore:AtomOpticsRealizationQKR:PRL95,Fishman:LocDynAnders:PRL82,Casati:IncommFreqsQKR:PRL89,Lemarie:Anderson3D:PRA09,Manai:Anderson2DKR:PRL15}. 

Recently it has been shown that Anderson localization can also occur in the time domain in systems that are perturbed by a force fluctuating in time \cite{Sacha2015a,sacha_delande16,DelandeMoralesSacha}. This phenomenon is related to time crystals, i.e. systems which can spontaneously switch to periodic motion \cite{Wilczek2012,Li2012}. While it is not easy to find a system that spontaneously moves when it is prepared in the ground state, models of time crystals with the help of periodically driven systems are interesting and can reveal new phenomena \cite{Chernodub2013,Wilczek2013,Bruno2013,Wilczek2013a,Bruno2013a,Li2012a,Bruno2013b,Guo2013,Watanabe2015,Sacha2015,khemani16,else16,Keyserlingk16,yao17,Weidinger16,zhang16,choi16,Syrwid17}. If a periodic perturbation starts fluctuating in time, then, in analogy to space crystals with disorder, Anderson localization of a system in the time domain can be observed. In other words, a detector situated at a certain position clicks with a probability that is localized exponentially around a certain moment in time. Anderson localization in the time domain is a general phenomenon that can be observed in many different systems where periodic driving starts behaving randomly \cite{sacha_delande16}. Despite the fact the time degree of freedom forms a one-dimensional {\it space}, it is possible to realize Anderson localization in the time domain with properties of multi-dimensional disorder systems where phase transition between localized and de-localized states can be observed \cite{DelandeMoralesSacha}. 

In the present paper we investigate Anderson localization in the time domain in a Rydberg atom perturbed by a fluctuating microwave field. Rydberg atom driven by a monochromatic microwave field can reveal non-spreading wave-packet motion where a resonantly driven electron is represented by a localized wave-packet that moves periodically along a classical orbit and does not spread. This phenomenon, discovered some time ago \cite{zas,hen+92,hol95,del+94,ibb,buch+95,dzb95,zdb95,sacha98,delande02} and observed in experiments \cite{maeda04,maeda07,maeda09,wyker12}, can be realized in different dynamical systems if non-linear classical resonances can form \cite{Buchleitner2002}. Switching from a monochromatic microwave field to a randomly fluctuating field we show that, in the frame moving with an electron along a classical orbit, Anderson localization takes place. In the laboratory frame, probability for detection of an electron at a given point on the orbit is localized exponentially around a certain moment of time. Such a behavior is repeated periodically in analogy to Anderson localization of a particle in a space crystal with disorder and with periodic boundary conditions where by traveling periodically around the ring, one observes periodically an exponentially localized density profile.

The paper is organized as follows. In  Sec.~\ref{s1d} we explain the phenomenon of the Anderson localization in the time domain in a one-dimensional (1D) model of an Hydrogen atom. Then, in Sec.~\ref{s3d} we switch to the 3D case and analyze realistic description of the system. Sec.~\ref{conl} is devoted to conclusions. 

\section{1D model}
\label{s1d}

Let us consider an Hydrogen atom perturbed by a linearly polarized microwave field in the 1D model \cite{Buchleitner2002}. The classical Hamiltonian of the system, in the dipole approximation and in the atomic units, reads 
\bea
H&=&H_0+H_1, 
\label{hinit0}
\\
H_0&=&\frac{p^2}{2}-\frac{1}{z}, \\ 
H_1&=&Fzf(t),
\label{hinit}
\eea
where $z\ge0$. Such a 1D model can describe classical elliptical orbits degenerated to a line along the field polarization axis. $F$ is a parameter that allows one to change the amplitude of the perturbation and $f(t)$ is a periodic function which describes time dependance of the electric field. The electric field is assumed to oscillate with a frequency $\omega$ but between $t=0$ and $t=2\pi/\omega$ it performs random fluctuations, i.e., 
\be
f(t)=f(t+2\pi/\omega)=\sum_{k\ne 0} f_k e^{ik\omega t},
\label{ft}
\ee
where Fourier components $f_k=f_{-k}^*$ are random numbers.

We are interested in a resonant driving of the atom when the frequency $\omega$ matches the frequency of electron motion on an unperturbed orbit. The description of the system close to the resonant trajectory can be significantly simplified if we employ the secular perturbation theory \cite{Buchleitner2002,Lichtenberg_s}. First, we switch to the action-angle variables, $J$ and $\theta$, of the unperturbed H atom that results in the following form of the Hamiltonian
\bea
H_0&=&-\frac{1}{2J^2}, \\
H_1&=&FJ^2\left(\frac{3}{2}-\sum_{m\ne 0}\frac{{\cal J}_m'(m)}{m}e^{im\theta}\right)
f(t),
\label{h1}
\eea
where ${\cal J}_m'$'s are the derivatives of the ordinary Bessel functions. The resonant condition means that
\be
\omega=\frac{\partial H_0(J_0)}{\partial J_0}=\frac{1}{J_0^3}.
\ee
In the moving frame, the position of an electron $\Theta=\theta-\omega t$ is a slow variable if we choose the conjugate momentum close to the resonant value $P=J-J_0\approx 0$. Then, averaging the Hamiltonian over time leads to the effective time-independent Hamiltonian
\be
H_{\rm eff}=-\frac{3}{2J_0^4}P^2+FJ_0^2\sum_{k\ne 0}\frac{{\cal J}_k'(k)}{k}f_{-k}e^{ik\Theta},
\label{heff}
\ee
where the constant term $-\frac{3}{2J_0^2}$ has been omitted. The first order secular approximation (\ref{heff}) is the accurate description of the system provided the second order contribution (that scales like $F^2J_0^6$) is small as compared to the first order term \cite{Lichtenberg_s}. It implies we may restrict ourselves to the first order effective Hamiltonian if $FJ_0^4\ll 1$ --- for discussion of the validity of the secular approximation in the context of the Anderson localization in time see appendix in \cite{sacha_delande16}.

Before we move on, we would like to discuss the secular Hamiltonian (\ref{heff}) and conditions necessary to observe Anderson localization we are interested in. If $f(t)$ was a simple single harmonic function like $\cos(k\omega t)$, then we would deal with a monochromatic resonant driving and well known non-spreading wave-packets can be realized \cite{Buchleitner2002}. That is, for $k=1$ we would deal with a single localized wave-packet moving along a Kepler orbit with the period $2\pi/\omega$ --- classically, in the laboratory frame, an electron would move along a Kepler orbit with small oscillations around it if the initial momentum $P\approx 0$. For $k>1$, superposition of $k$ such wave-packets would move along a Kepler orbit --- each wave-packet evolves with a period $k2\pi/\omega$ but each of them is delayed with respect to its neighbor by $2\pi/\omega$ \cite{Buchleitner2002}. We will see that in order to observe the Anderson localization we need $f(t)$ to consist of many harmonics at the same time. Then, the secular Hamiltonian (\ref{heff}) is the result of the coherent addition of resonant terms between the spatial harmonics of an unperturbed electronic motion and the corresponding temporal harmonics of the disordered driving amplitude. It also indicates that the electric dipole moment $z$ expressed in terms of the action-angle variables must possess many spatial harmonics, cf. Eq.~(\ref{h1}). It is possible provided an electron moves along an elliptical orbit and approaches closely to the nucleus. Thus, circular Kepler orbits are not suitable for realization of the Anderson localization in the time domain.

Harmonics of an unperturbed motion of an electron decrease with $k$ like ${\cal J}'_k(k)/k$, cf. Eq.~(\ref{h1}). If the fluctuations of the microwave field are engineered so that the components $f_{-k}$ fulfill
\be
\left|\frac{{\cal J}_k'(k)}{k}f_{-k}\right|=\frac{1}{\sqrt{2k_0}},
\label{fkcond}
\ee
for $|k|\le k_0$ and zero otherwise, and $\varphi_{-k}=-\varphi_{k}={\rm Arg}(f_{-k})$ are random numbers chosen uniformly from the interval $[0,2\pi)$, we end up with the effective Hamiltonian,
\bea
H_{\rm eff}&=&\frac{P^2}{2m_{\rm eff}}+FJ_0^2V(\Theta), 
\label{heffs}
\\
V(\Theta)&=&\frac{1}{\sqrt{2k_0}}\sum_{\substack{|k|\le k_0\\k\ne 0}}e^{i(k\Theta+\varphi_k)}, 
\label{vdis} 
\\
m_{\rm eff}&=&-\frac{J_0^4}{3},
\eea
that describes a particle with the negative effective mass $m_{\rm eff}$ in the disorder potential characterized by the variance $F^2J_0^4$ and the finite correlation length $\zeta=\sqrt{2}/k_0$ \cite{sacha_delande16}. Note, that $F$ is not the amplitude of the electric field because the microwave field is not monochromatic and $\frac{\omega}{2\pi}\int_0^{2\pi/\omega}dt f^2(t)\ne \frac12$. The intensity of the microwave field reads
\be
I=\frac{F^2}{2} \sum_{k\ne 0}|f_k|^2=\frac{F^2}{4k_0}\sum_{\substack{|k|\le k_0\\k\ne 0}}\left(\frac{k}{{\cal J}'_k(k)}\right)^2.
\label{intensity}
\ee
In the present paper we have chosen the fluctuations of the microwave field that lead to the disorder potential (\ref{vdis}) as an example. By a proper choice of $f(t)$ one can realize many different effective disorder potentials. 

In the case of the quantum version of the Hamiltonian (\ref{heffs}), i.e. quantizing the action angle variables (where $J_0$ becomes the principal quantum number $n_0$ of an H atom), we can expect Anderson localization. In 1D systems the presence of a disorder results in Anderson localization regardless how weak the disorder is. In the present case, the configuration space extends from $\Theta=0$ to $2\pi$ due to the periodic boundary conditions. Therefore, Anderson localization can be observed if the localization length $\xi_{\rm loc}\ll 2\pi$. In the weak disorder limit, i.e. when $F^2n_0^4\ll EE_\zeta$ where $E$ is an energy eigenvalue and $E_\zeta=k_0^2/2m_{\rm eff}$ is the so-called correlation energy, $\xi_{\rm loc}$ can be calculated by means of the Born approximation \cite{MuellerDelande:Houches:2009,Kuhn:Speckle:NJP07} and it reads
\be
\frac{\xi_{\rm loc}}{\zeta}=\frac{2\sqrt{2}}{\pi}\frac{EE_\zeta}{F^2n_0^4},
\label{born}
\ee
for $E\ge E_\zeta/4
$.
Note, that the effective mass $m_{\rm eff}$ is negative, thus, the Hamiltonian (\ref{heffs}) is bounded from above not from below as in a usual case. Consequently, the greatest eigenvalue of (\ref{heffs}) corresponds to the strongest localization. The localization length increases with a decrease of $E$. The validity of the Born approximation requires the localization length $\xi_{\rm loc}$ to be much greater than the correlation length $\zeta$ that, together with the requirement $\xi_{\rm loc}\ll 2\pi$, can be easily fulfilled for sufficiently large $n_0$ and with an appropriate choice of $k_0$.


\begin{figure}
\includegraphics[width=0.49\columnwidth]{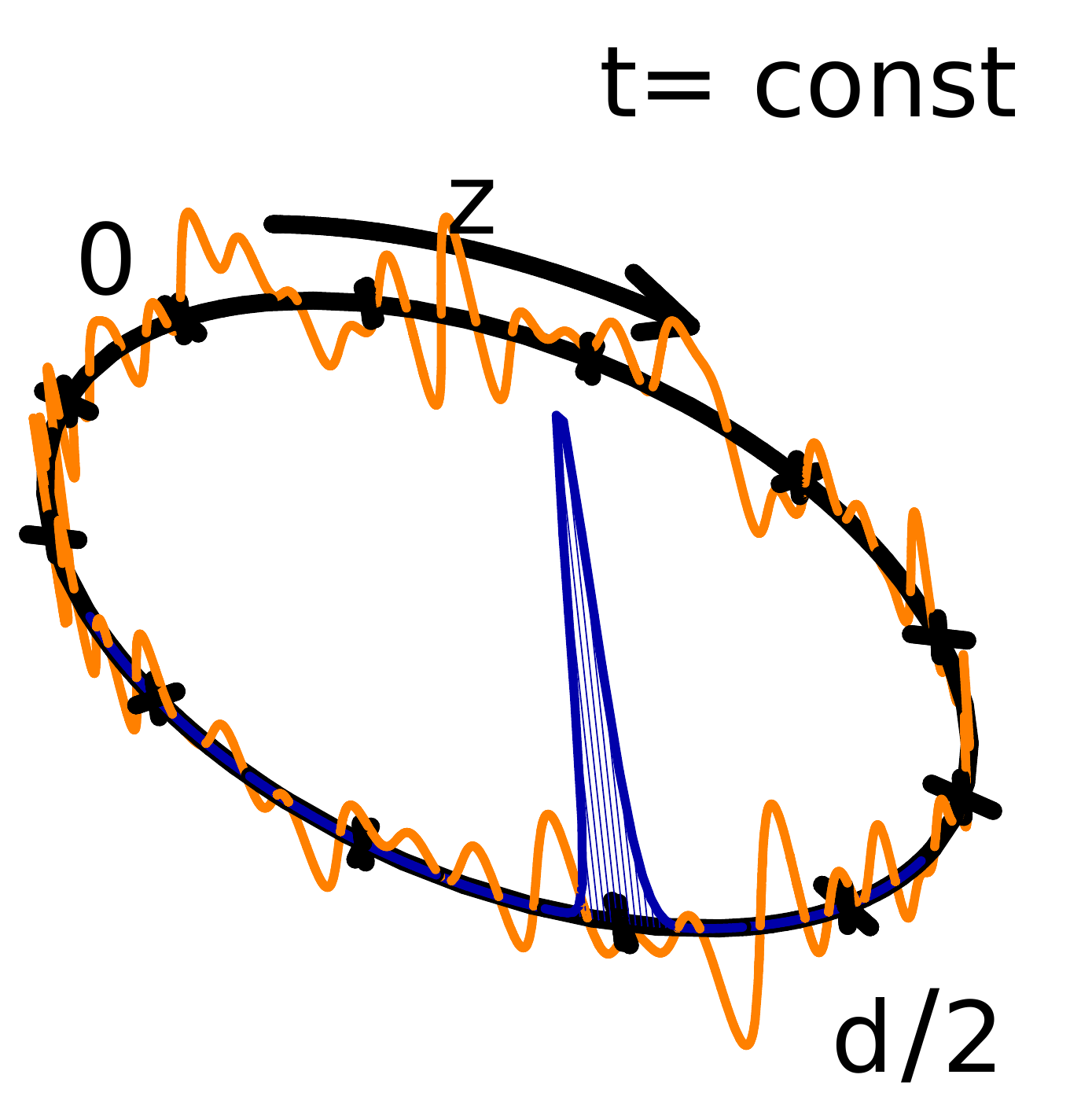}
\includegraphics[width=0.49\columnwidth]{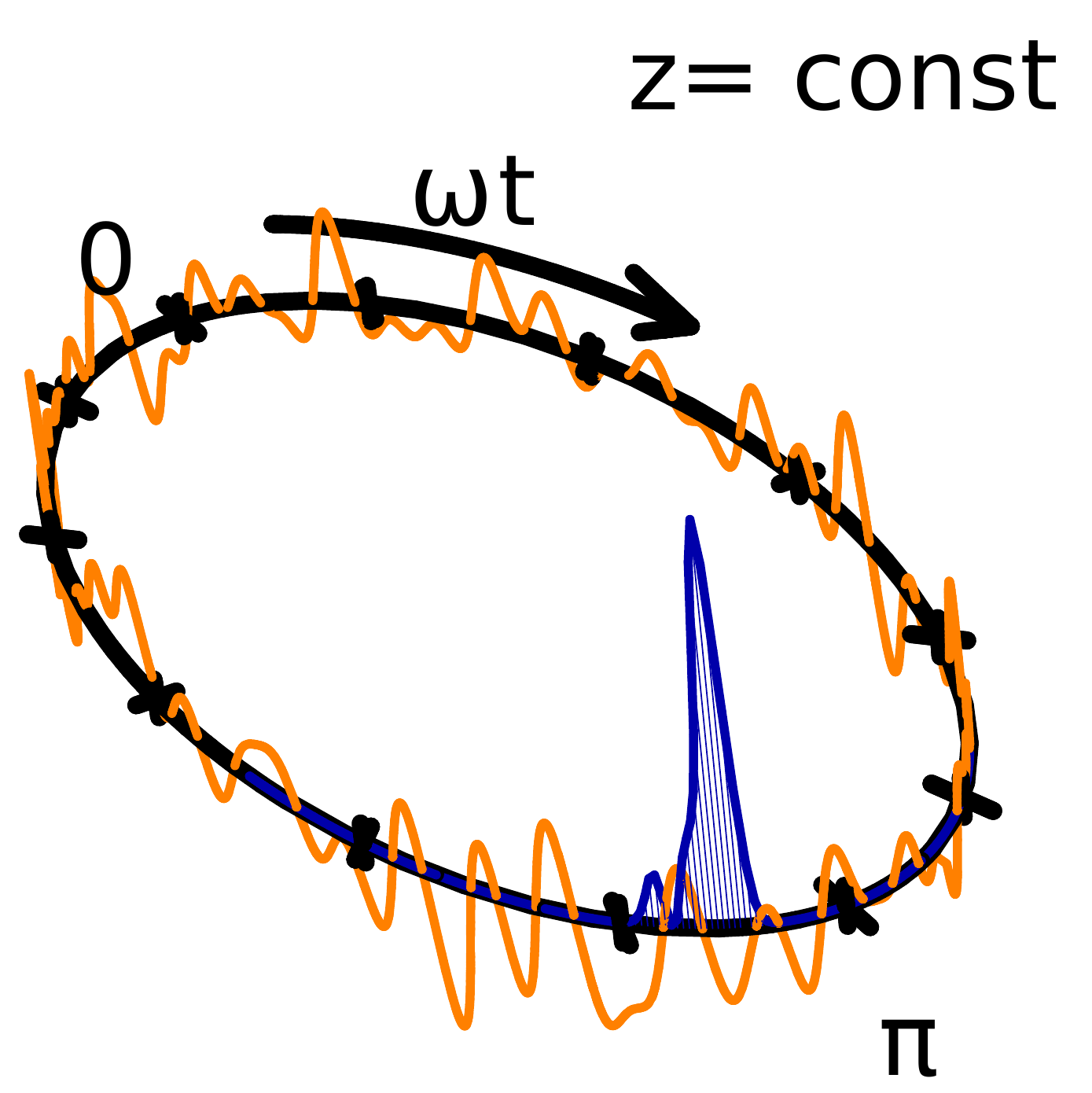}
\caption{(color online) Left panel: schematic illustration of the standard Anderson localization in the configuration space, i.e. localization of a particle on a ring of length $d$ in the presence of a time-independent disordered potential with periodic boundary conditions. Right panel illustrates Anderson localization in the time domain --- for a fixed position in the configuration space, the probability density for the detection of a particle is exponentially localized around a certain moment of time. Such a behavior is repeated with a period $T=2\pi/\omega$ similarly like in the left panel where if one travels periodically around the ring, one observes periodically an exponentially localized density profile.}
\label{schematic}
\end{figure}

Anderson localization in the moving frame that is predicted with the help of the Hamiltonian (\ref{heffs}) means an exponential localization in the time domain when we switch to the laboratory frame \cite{Sacha2015a,sacha_delande16,DelandeMoralesSacha}. Indeed, in the laboratory frame a localized wavepacket will move along an unperturbed classical orbit. If we ask how the probability for the detection of an electron, at a fixed position on an orbit, changes in time, it will turn out that the exponential localization in the $\Theta$ space translates into an exponential localization in time around a certain $t_0$ because for a fixed position in the laboratory frame the relation between $\Theta$ and $t$ is linear, i.e. $\Theta=\theta-\omega t$. However, if we fix $t$ and ask whether an electron is also exponentially localized in the configuration space, the answer is no because the transformation from the $\Theta$ space to the Cartesian space is nonlinear, cf. (\ref{hinit}) and (\ref{h1}). In Fig.~\ref{schematic} we illustrate schematically the idea of Anderson localization in the time domain and compare it to the standard Anderson localization in the configuration space in systems with spatially disordered potentials and periodic boundary conditions.

\begin{figure}
\includegraphics[width=1.\columnwidth]{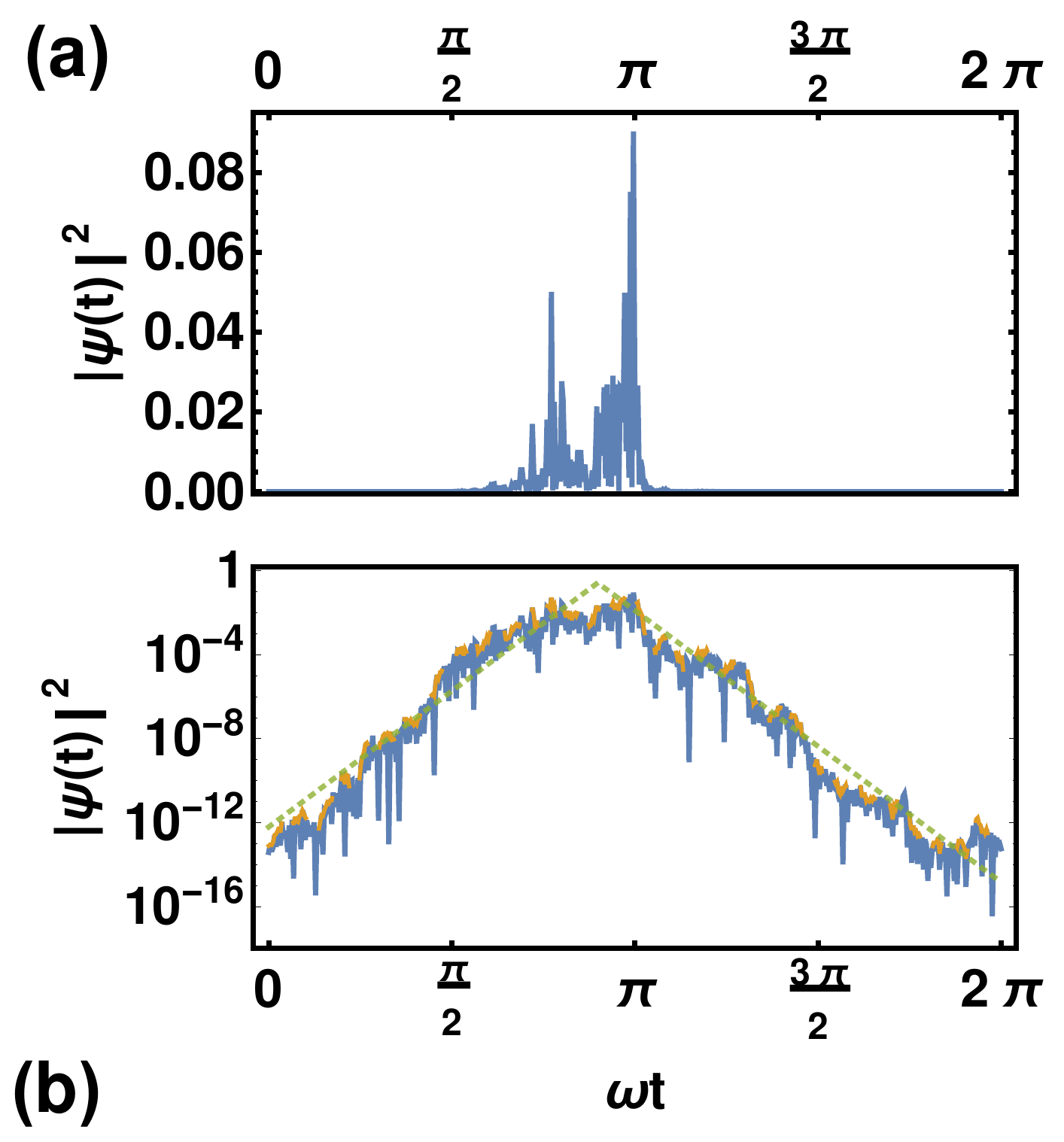}
\caption{(color online) Hydrogen atom in a fluctuating microwave field in the 1D model. Panels show probability density for the detection of an electron at the nucleus versus time (solid blue lines) in linear (a) and logarithmic (b) scale corresponding to a Floquet eigenstate with a quasi-energy $n_0^2E=-\frac{3}{2}-1.53\cdot 10^{-8}$ of the Hamiltonian (\ref{qheff}). Microwave field is resonant with electron motion, i.e. $\omega=n_0^{-3}$ where $n_0=10^6$, and consists of $k_0=500$ harmonics with randomly chosen relative phases; the amplitude parameter $Fn_0^4=1.6\times 10^{-8}$. An additional smooth envelope [orange dashed line in (b)] corresponds to the norm $||\psi(t)||^2=|\psi(t)|^2+\alpha|\partial_t\psi(t)|^2$ which better visualizes an exponential profile --- small positive $\alpha$ is introduced to avoid zeros of $||\psi(t)||$ \cite{lugan}. For such system parameters, the localization length in the time domain can be predicted with the help of the Born approximation (\ref{born}) which results in $0.18/\omega$
 that agrees quite well with the value of the localization length, $0.21/\omega$, obtained by a numerical fit of an exponentially localized function [dotted green line in (b)].}
\label{prob1D}
\end{figure}

\begin{figure}
\includegraphics[width=1.\columnwidth]{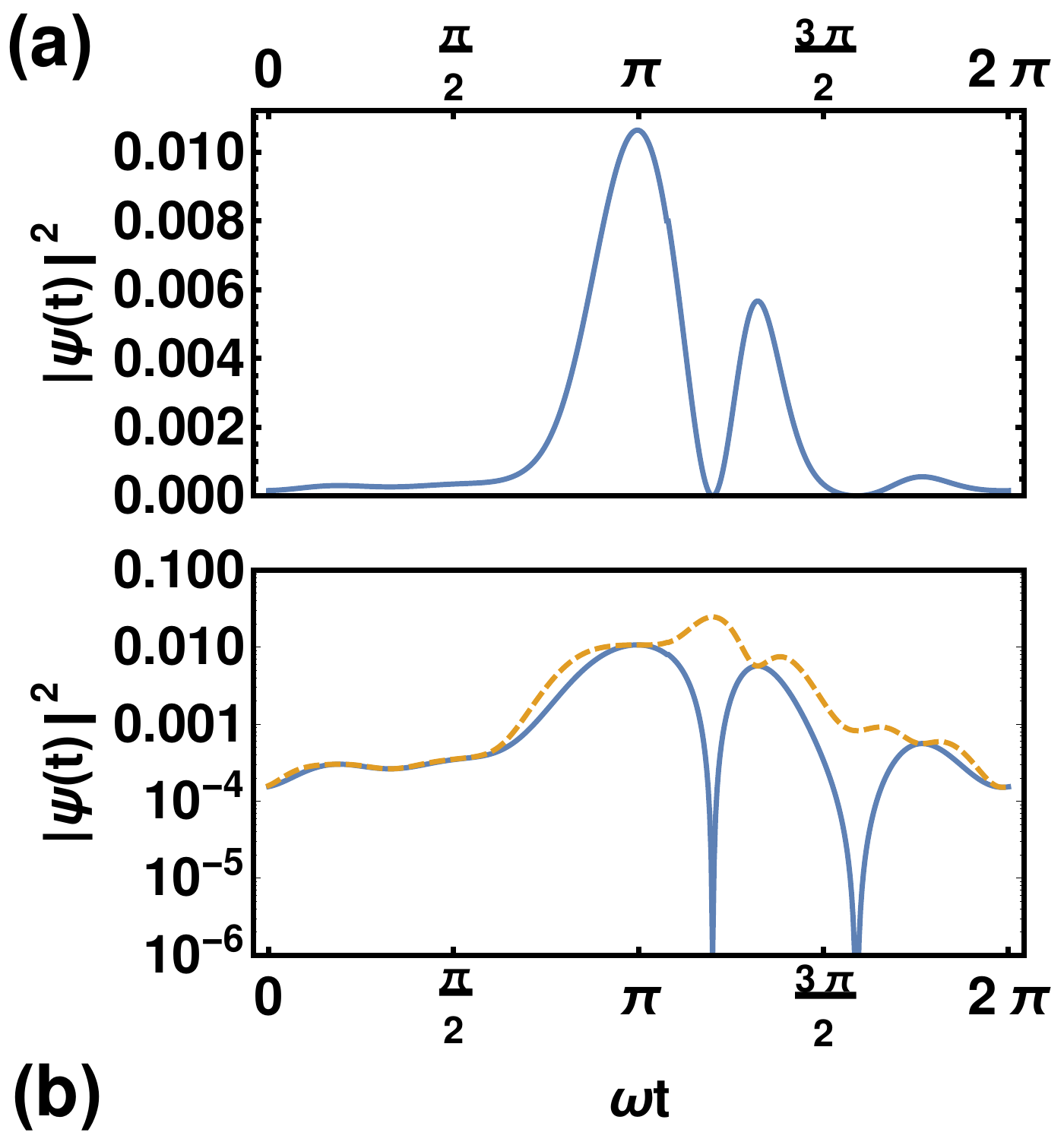}
\caption{(color online) The same as in Fig.~\ref{prob1D} but for parameters attainable in a laboratory, i.e. $\omega=n_0^{-3}$ where $n_0=300$, $k_0=5$ and $Fn_0^4=0.00016$ [i.e. the intensity $I=0.72~\frac{W}{cm^2}$, Eq.~(\ref{intensity})]. The Floquet eigenstate corresponding to the third (from above) quasi-energy of the Hamiltonian (\ref{qheff}) is chosen, i.e. $n_0^2E=-\frac{3}{2}-2.45\cdot 10^{-5}$.}
\label{prob1Da}
\end{figure}

We have obtained the prediction for Anderson localization in the time domain by means of the effective classical Hamiltonian and subsequent quantization of the classical action angle variables. However, the same results can be obtained starting with the full quantum Hamiltonian and applying a quantum version of the secular approach \cite{zas}. The Hamiltonian (\ref{hinit0}) is time periodic which means that although there are no energy eigenstates, one can find a kind of {\it stationary} states $\psi_m(z,t)$ that are time periodic eigenfunctions of the so-called Floquet Hamiltonian $H_F(t)$ \cite{Buchleitner2002},
\be
H_F\psi_m=(H-i\partial_t)\psi_m=E_m\psi_m,
\ee
where $E_m$ are real eigenvalues which are called quasi-energies of a system. The Floquet theorem \cite{shirley65}, that is used here, is in a full analogy to the Bloch theorem known in condensed matter physics but we deal with a system periodic in time not in space. 

Performing time-dependent unitary transformation $\hat U=e^{i\hat n\omega t}$, where $\hat n=(-2\hat H_0)^{-2}$ is the operator of the principal quantum number of an Hydrogen atom, we switch to the moving frame similarly as in the classical approach. Then, the matrix elements of the resulting Floquet Hamiltonian in the hydrogenic eigenbasis read,
\bea
\langle n'|\hat H_F'(t)|n\rangle&=&\langle n'|\hat U\hat H_F(t)\hat U^\dagger|n\rangle
\cr
&=&\left(-\frac{1}{2n^2}-n\omega-i\partial_t\right)\delta_{nn'}
\cr
&& +F\langle n'|z|n\rangle f(t)e^{-i(n-n')\omega t}.
\label{hprim}
\eea
Averaging the last term of (\ref{hprim}) over time we obtain the quantum version of the effective Hamiltonian \cite{zas},
\be
\langle n'|\hat H_{\rm eff}|n\rangle=\left(-\frac{1}{2n^2}-n\omega\right)\delta_{nn'}+F\langle n'|z|n\rangle f_{n-n'},
\label{qheff}
\ee
where the operator $i\partial_t$ has been omitted because for the time-independent effective Hamiltonian it would only introduce a shift of energy eigenvalues. The effective Hamiltonian (\ref{qheff}) corresponds to a single block of the matrix of the Floquet Hamiltonian $\langle n',k'|\hat H'_{F}|n,k\rangle$ in the basis spanned by the hydrogenic eigenvectors $|n\rangle$ and the Fourier functions $\langle t|k\rangle=\sqrt{\frac{\omega}{2\pi}}e^{ik\omega t}$, i.e. to the block with $k=k'=0$. The omission of couplings between the chosen block and other blocks is valid if we want to describe the Hilbert subspace of states with principal quantum numbers $n$ close to the resonant value $n_0$, where $n_0=\omega^{-1/3}$, and if $Fn_0^4\ll 1$. 

Numerical diagonalization of (\ref{qheff}) allows us to obtain Floquet eigenstates $\psi_m'(z)$ of the system in the moving frame within the secular approximation. Switching to the laboratory frame, $\psi_m(z,t)=e^{-i\hat n\omega t}\psi_m'(z)$, we expect that time evolution of the probability density for the detection of an electron at a fixed position will reveal exponential localization in time if Anderson localization takes place. 

In Figs.~\ref{prob1D}-\ref{prob1Da} we show how the probability for the detection of an electron at the position of the nucleus of an H atom changes in time for two different sets of the parameters. Figure~\ref{prob1D} presents the results for the case of $n_0=10^6$ where the Anderson localization can be described with the help of the Born approximation while in Fig.~\ref{prob1Da} we have used values of the parameters that are attainable in present-day laboratories. In order to afford small correlation length of the effective disordered potential, $\zeta=\sqrt{2}/k_0$, a big value of $k_0$ is needed. Then, however, the value of $F$ must be decreased to stay in the validity range of the effective Hamiltonian because the intensity (\ref{intensity}) increases with $k_0$. As the result, high excitation of a Hydrogen atom is necessary in order to observe the Anderson localization in time. In Fig.~\ref{prob1Da} we have chosen an experimentally attainable value $n_0=300$. Then for $k_0\approx 5$ and suitable choice of $F$ the effective Hamiltonian is valid and we can see signatures of Anderson localization in the time domain.

We would like to note that the spectra obtained in diagonalization of the quantum secular Hamiltonian (\ref{qheff}) and the quantized version of the classical secular Hamiltonian (\ref{heffs}) are identical and the same is true for the time evolution of the probabilities for the detection of an electron at the nucleus that are shown in Figs.~\ref{prob1D}-\ref{prob1Da}.

\section{3D description}
\label{s3d}

The 1D model allowed us to present the idea of Anderson localization in the time domain in a Rydberg atom and to explain the classical and quantum secular approximation methods. 
In this section we show that the Anderson localization predicted in Sec.~\ref{s1d} survives when we switch to the 3D description. Moreover, the 3D case allows also for observation of Anderson localization of an electron not only on an alongated one-dimensional orbit but also on an elliptical trajectory. 

We consider an H atom perturbed by a linearly polarized microwave field, similarly as in Sec.~\ref{s1d}, but now we add also a static electric field along the polarization axis of the microwave field, i.e. the Hamiltonian of the system reads $H=H_0+H_1$ with
\be
H_0=\frac{{\vect p}^2}{2}-\frac{1}{r}, \quad H_1=Fzf(t)-F_s z,
\label{h3d}
\ee
where $f(t)$ fulfills (\ref{ft}) and $F_s$ stands for the static electric field amplitude. Projection of the angular momentum of an electron on the $z$ axis is conserved and in the following we assume it is zero. 

\begin{figure}
\includegraphics[width=1.\columnwidth]{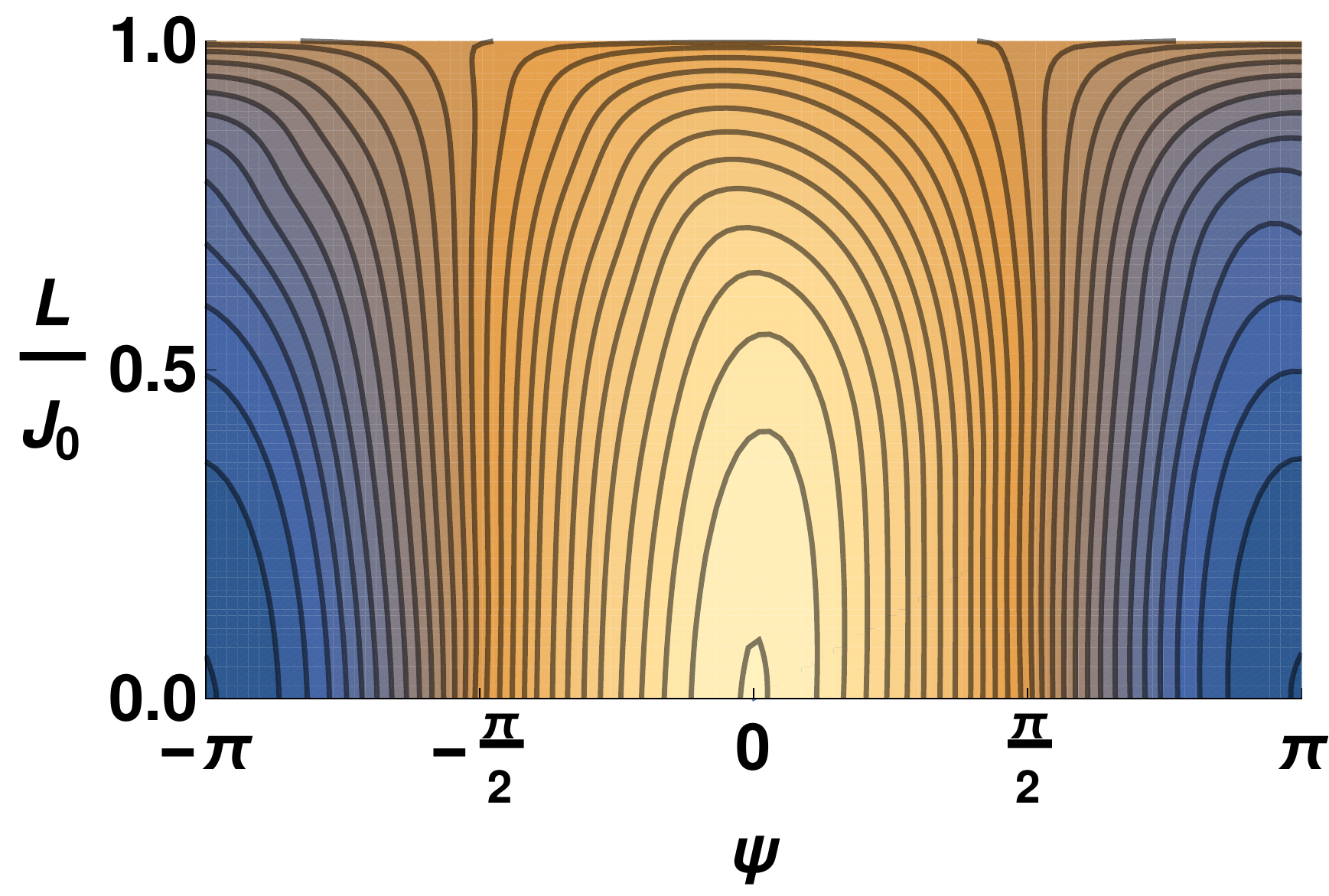}
\includegraphics[width=1.\columnwidth]{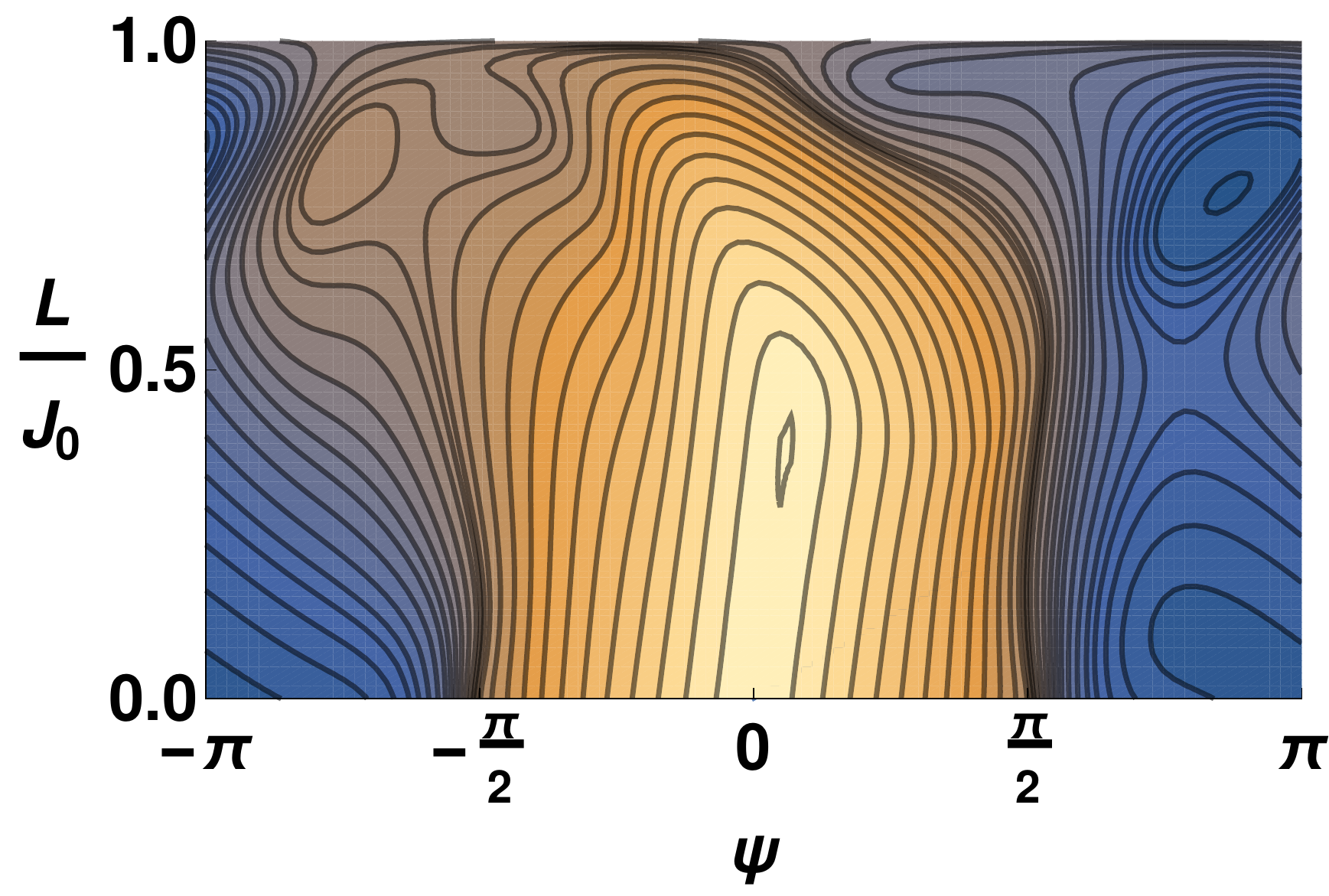}
\caption{(color online) Structure of the $(L,\Psi)$ phase space related to the effective Hamiltonian (\ref{heff3d}) after the first stage of the Born-Oppenheimer approximation, i.e. when the motion of $J$ and $\Theta$ was quantized and, e.g., the third eigenenergy (from above) was chosen. Then, the total energy is a function $L$ and $\Psi$ only, and its iso-value contours are presented for $F_s/F=1.5$ (top panel) and $F_s/F=0.33$ (bottom panel) --- the lighter area, the greater energy. Position $(L_0=0,\Psi_0=0)$ of the stable fixed point visible in the top panel depends neither on an  eigenenergy chosen in the first stage of the Born-Oppenheimer approach nor on a realization of the fluctuating microwave field. Position $(L_0/J_0=0.4,\Psi_0=0.06\pi)$ of the fixed point visible in the bottom panel changes weakly in different realizations of a randomly fluctuating field --- this fixed point corresponds to an elliptical Kepler orbit whose major axis is not precisely oriented along the $z$ axis. The other parameters are the following: $\omega=J_0^{-3}$ where $J_0=300$, $J_0^4F=0.00016$, $f_k$ like in (\ref{fkcond}) with $k_0=5$.
}
\label{lpsi}
\end{figure}

\begin{figure}
\includegraphics[width=1.\columnwidth]{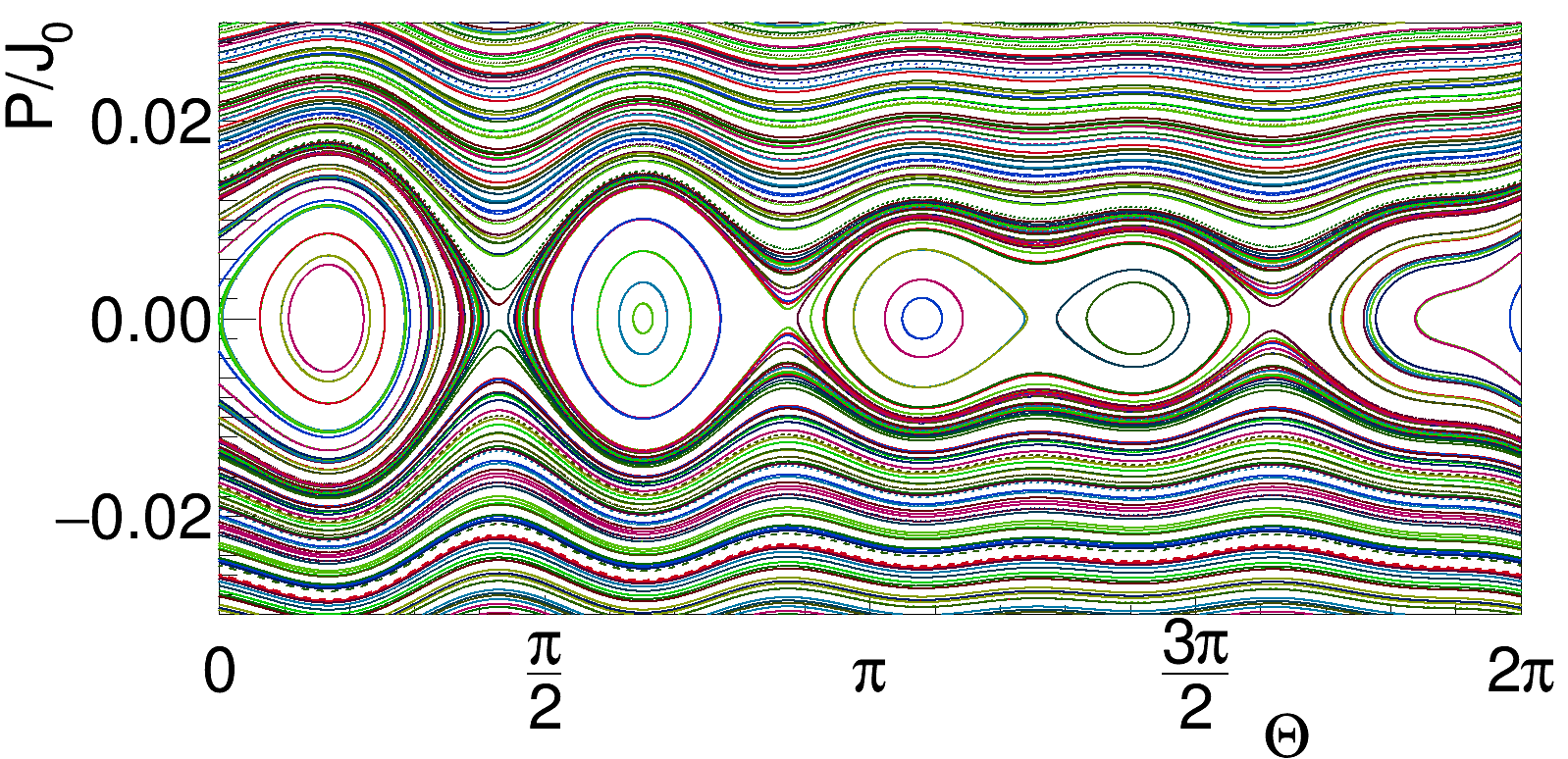}
\includegraphics[width=1.\columnwidth]{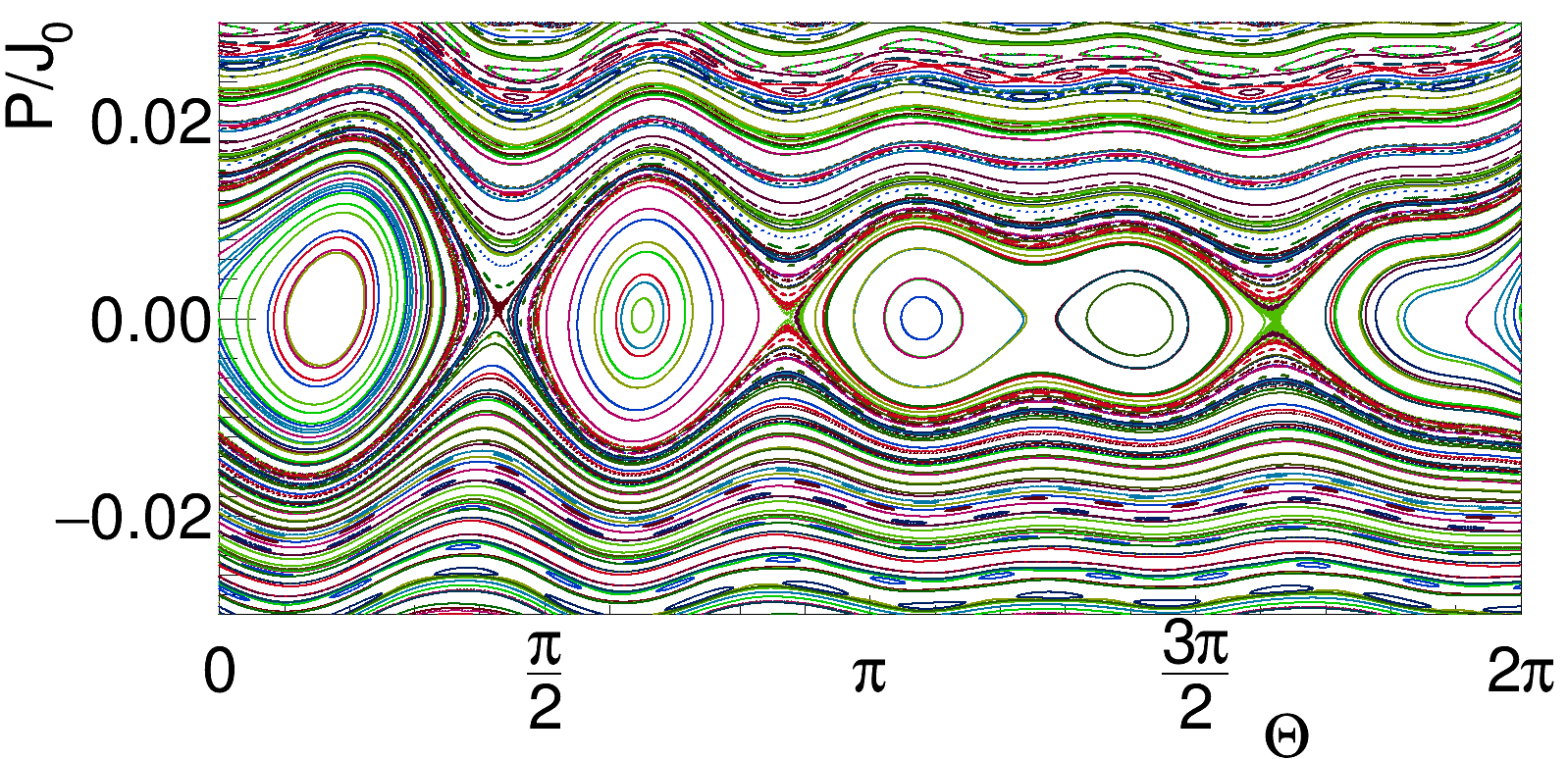}
\caption{(color online) Top panel: portrait of the $(P,\Theta)$ phase space generated by the 1D effective Hamiltonian (\ref{heffs}). Bottom panel: stroboscopic picture of the $(P,\Theta)$ phase space obtained in numerical integration of the full classical equations of motion with initial values: $L/J_0=0.1$ and $\Psi=0$. The values of the parameters are the following: $FJ_0^4=0.00016$, $F_sJ_0^4=0.00024$, $f_k$ like in (\ref{fkcond}) with $k_0=5$.}
\label{sos}
\end{figure}

\begin{figure}
\includegraphics[width=1.\columnwidth]{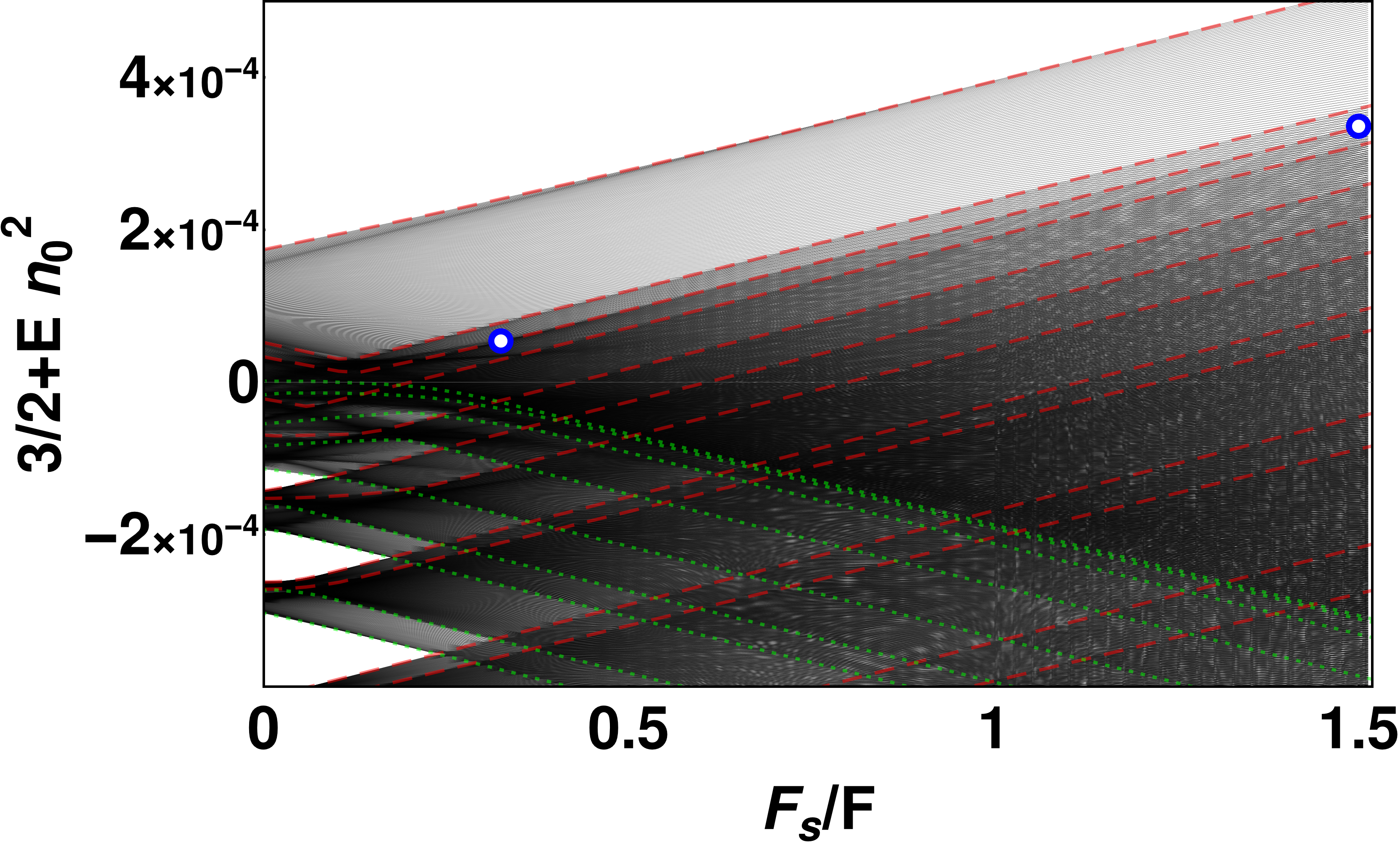}
\caption{(color online) Black solid lines: quasi-energy levels versus $F_s$ obtained in diagonalization of the 3D quantum effective Hamiltonian (\ref{qheff_3d}). Red dashed lines: eigenvalues obtained semiclassically and within the Born-Oppenheimer approximation that correspond to the stable fixed point of the highest energy visible in Fig.~\ref{lpsi} --- different dashed lines are related to different quantum numbers of the quantized $(P,\Theta)$ motion. Green dotted lines: similar semiclassical eigenvalues as indicated with red dashed lines but related to a fixed point in the $(L,\Psi)$ space of the lowest energy. Opens circles: indicate quasi-energy levels that correspond to two Floquet eigenstates presented in Figs.~\ref{3Dloc_in_time}-\ref{den_config}. The other parameters are the following: $\omega=n_0^{-3}$ where $n_0=300$, $Fn_0^4=0.00016$, $f_k$ like in (\ref{fkcond}) with $k_0=5$.}
\label{leveldyn}
\end{figure}

\begin{figure}
\includegraphics[width=1.\columnwidth]{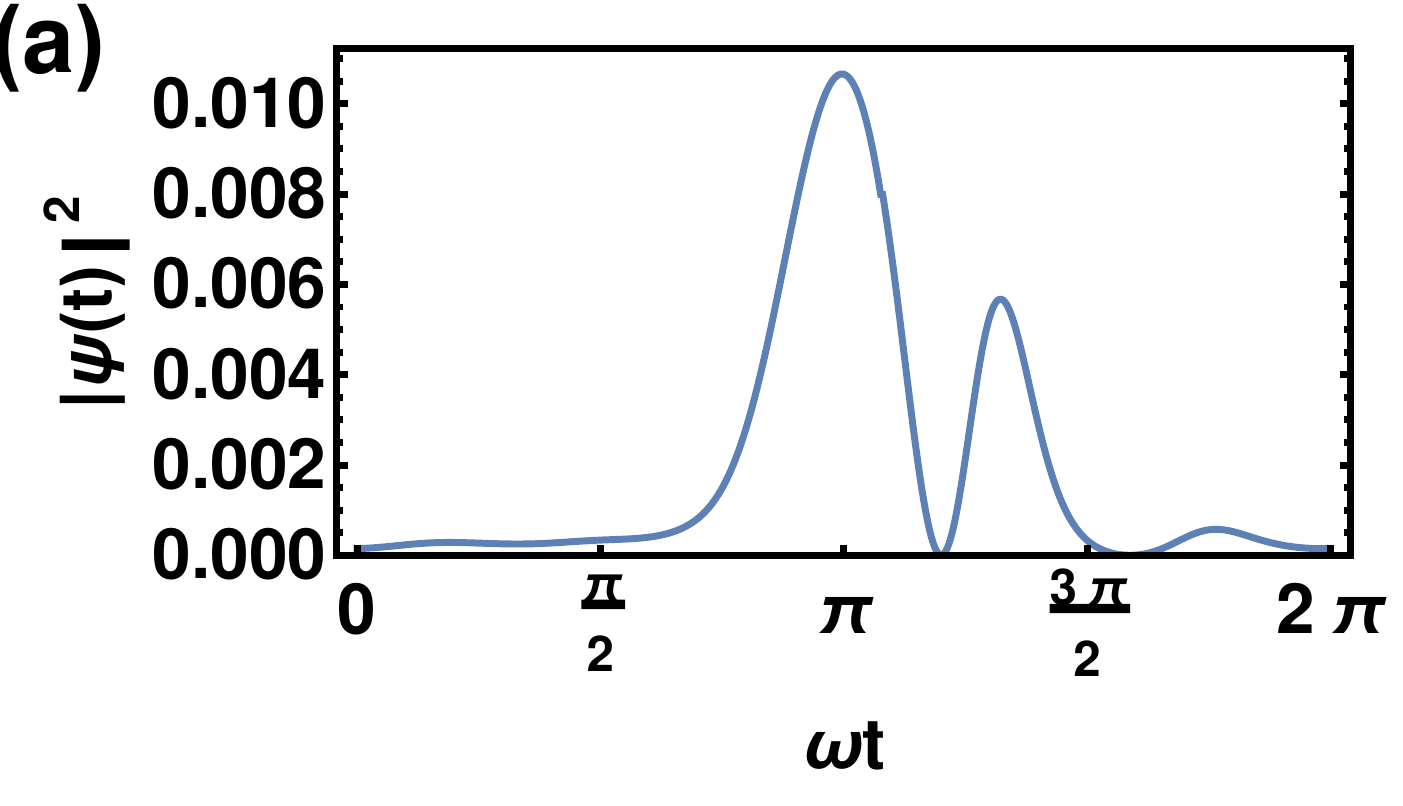}
\includegraphics[width=1.\columnwidth]{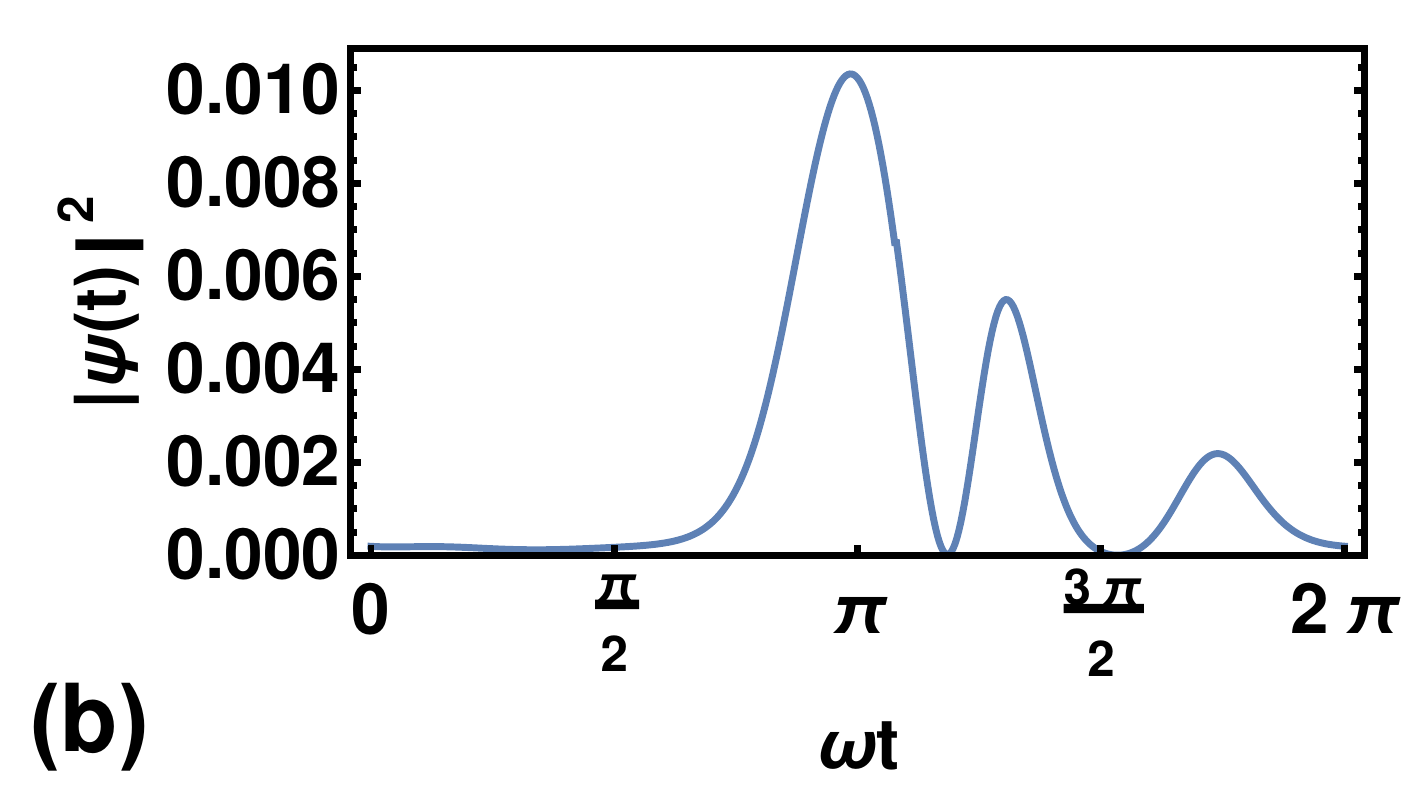}
\caption{Hydrogen atom in a fluctuating microwave field in the 3D model. Panels show probability density for the detection of an electron at the nucleus versus time for $F_s/F=1.5$ (a) and $F_s/F=0.33$ (b). The Floquet eigenstate presented in (a) corresponds to the eigenstate obtained in the 1D model and shown in Fig.~\ref{prob1Da}. Panel (b) presents a Floquet eigenstate related to Anderson localization of an electron on an elliptical trajectory. The corresponding quasi-energy levels are indicated in Fig.~\ref{leveldyn} with open circles. All parameters are the same as described in Fig.~\ref{leveldyn}.
}
\label{3Dloc_in_time}
\end{figure}

\begin{figure}
\includegraphics[width=0.295\columnwidth]{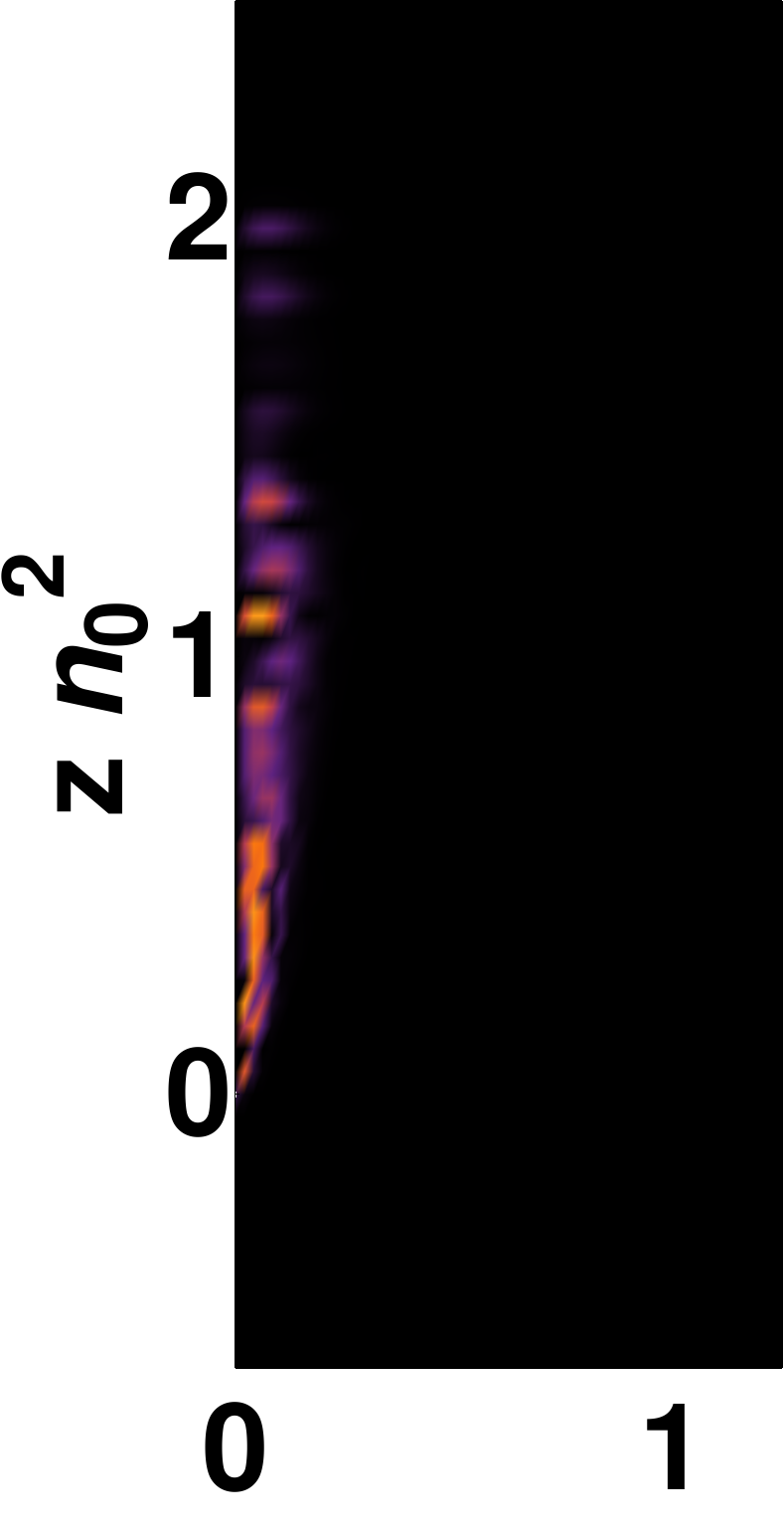}
\includegraphics[width=0.218\columnwidth]{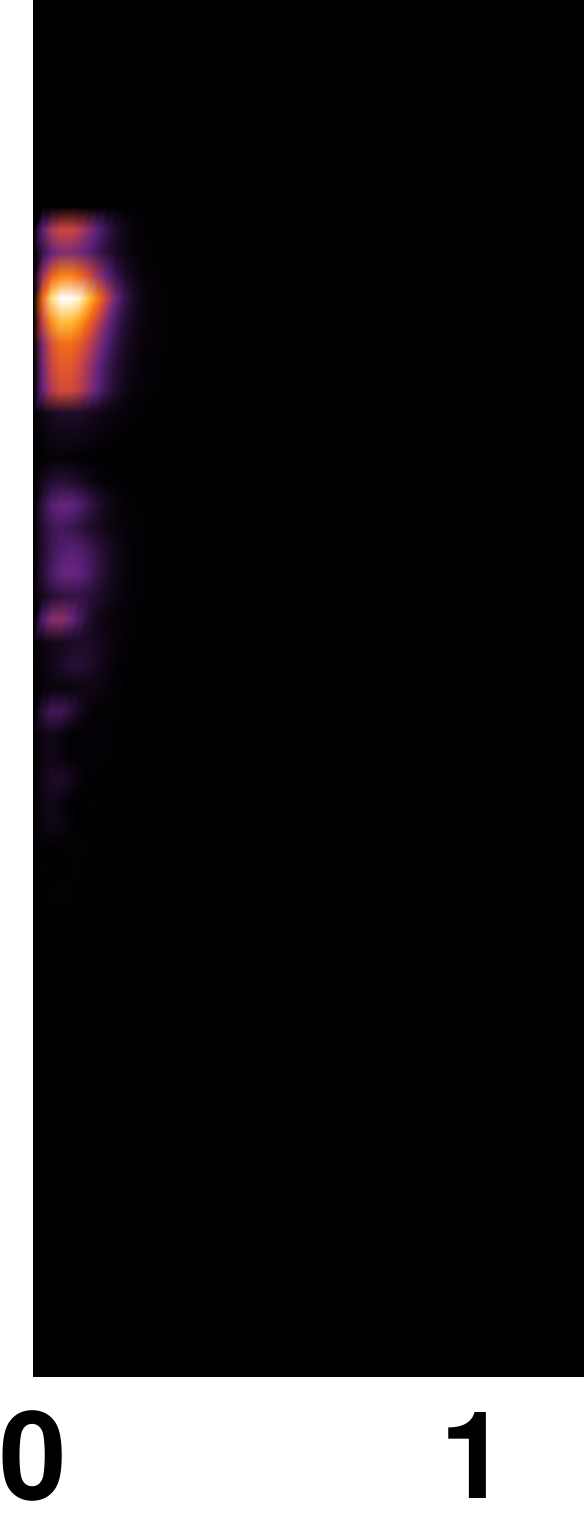}
\includegraphics[width=0.218\columnwidth]{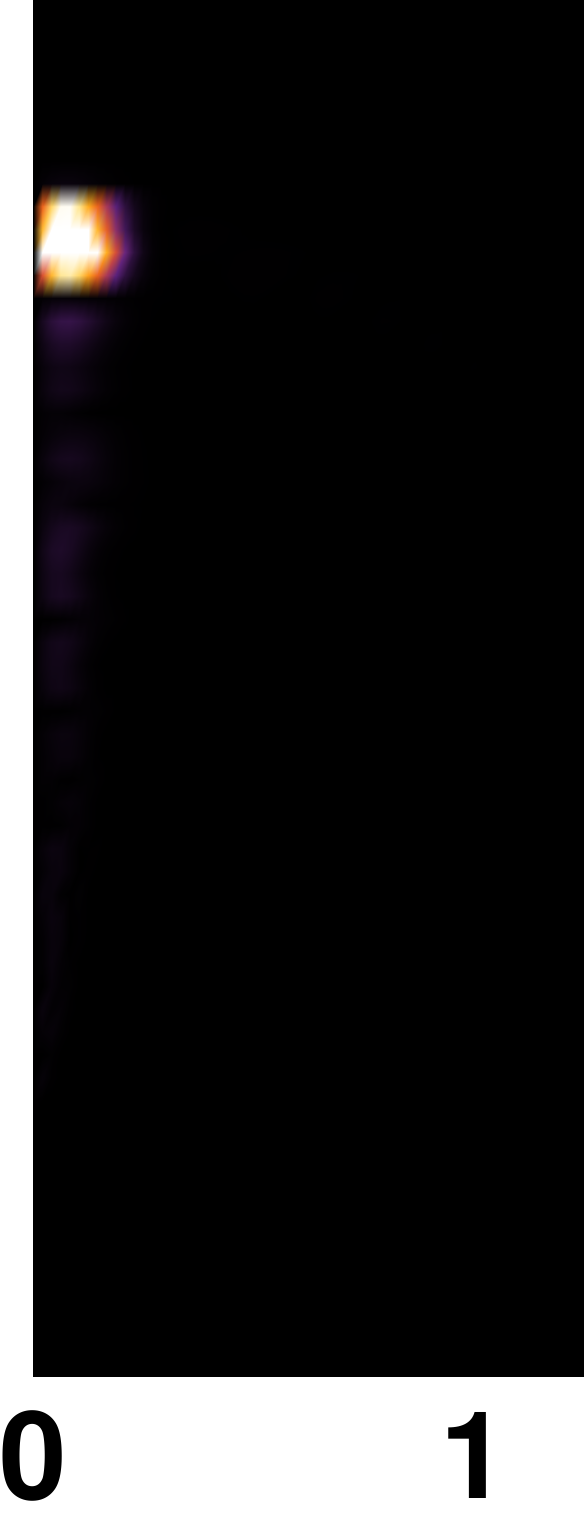}
\includegraphics[width=0.218\columnwidth]{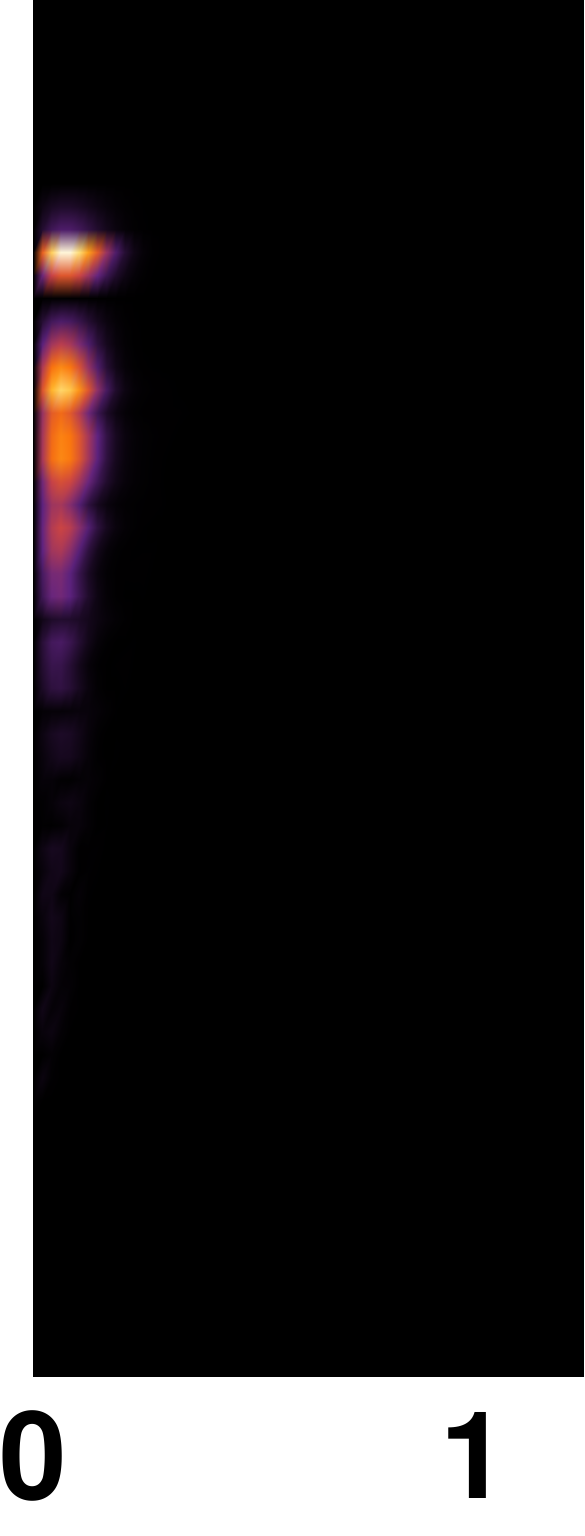}\\
\includegraphics[width=0.3\columnwidth]{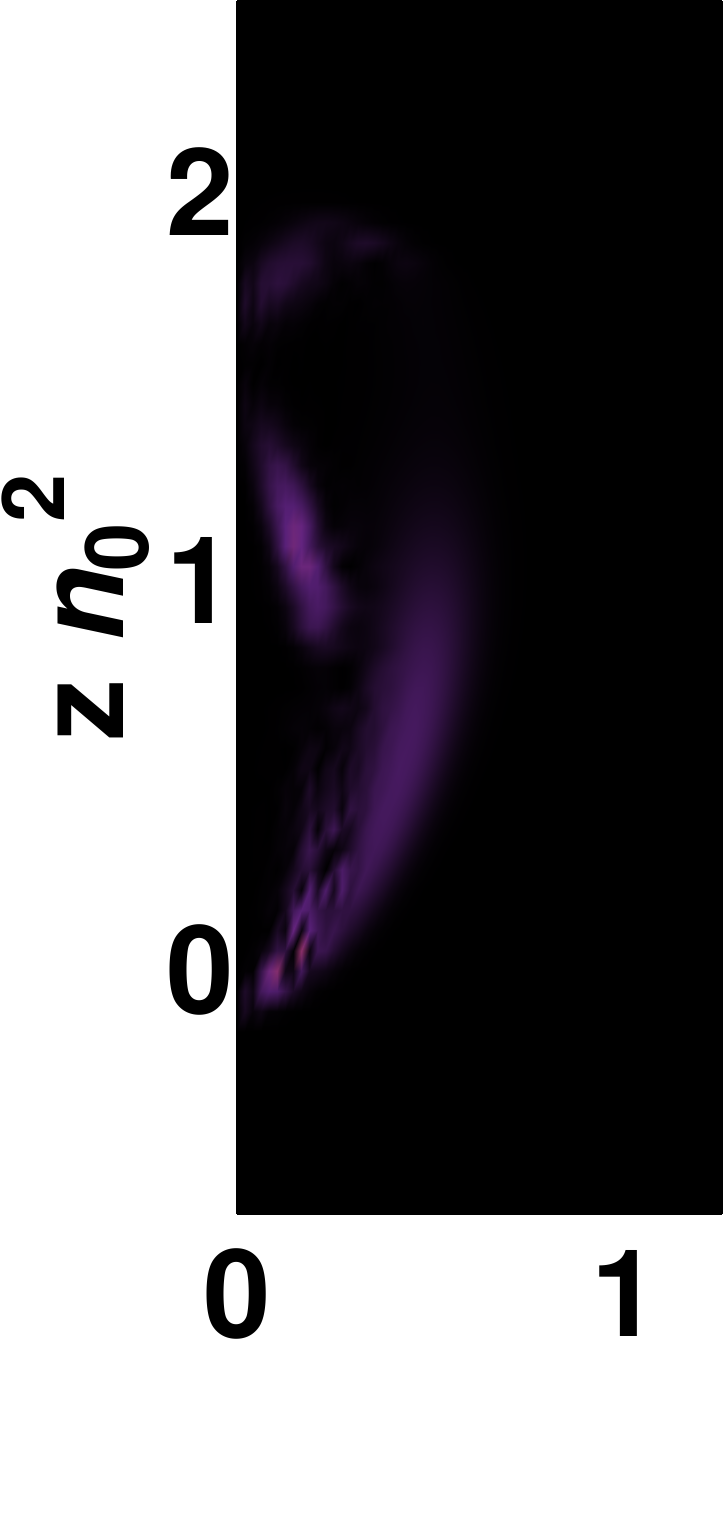}
\includegraphics[width=0.217\columnwidth]{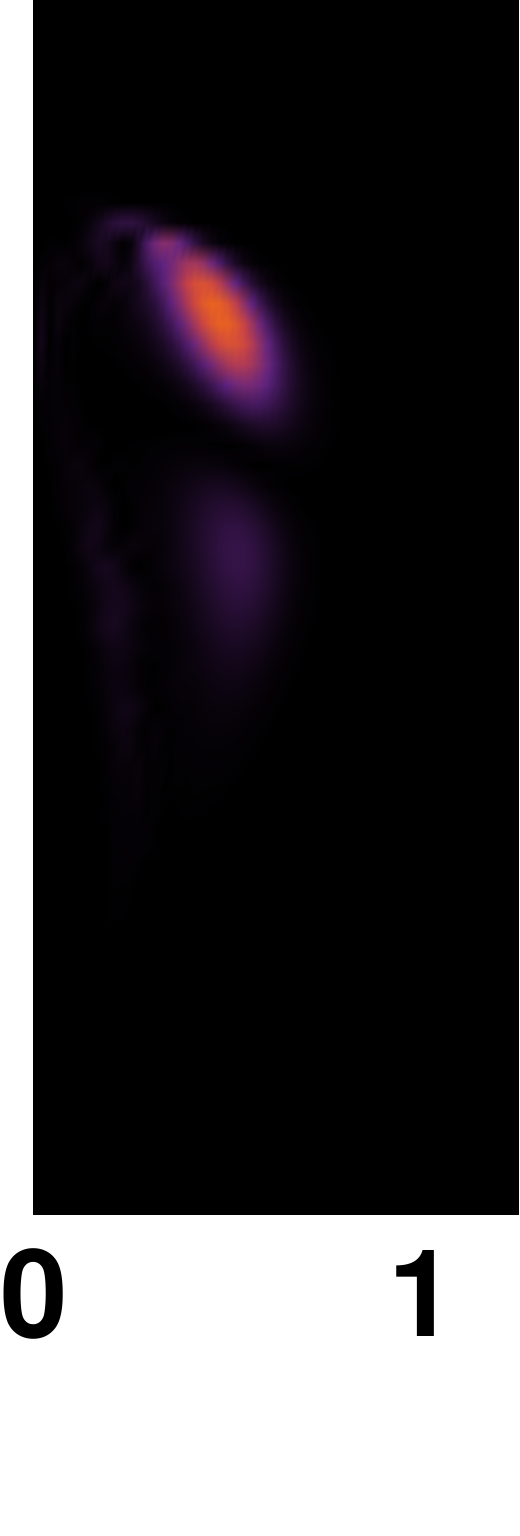}
\includegraphics[width=0.217\columnwidth]{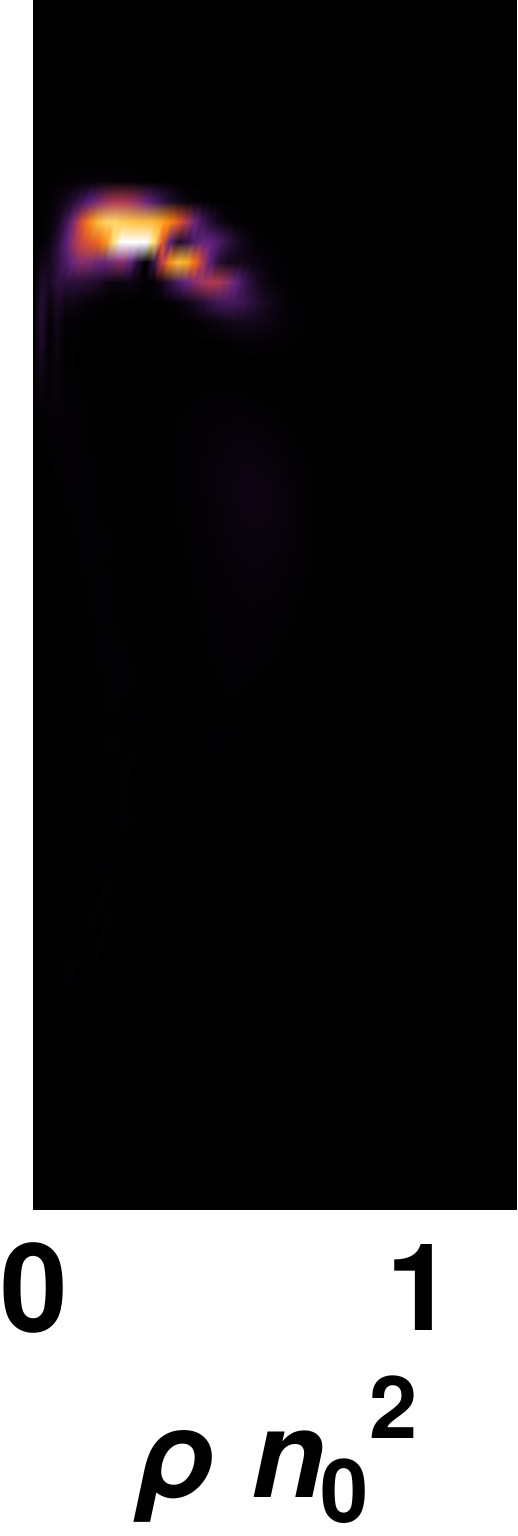}
\includegraphics[width=0.217\columnwidth]{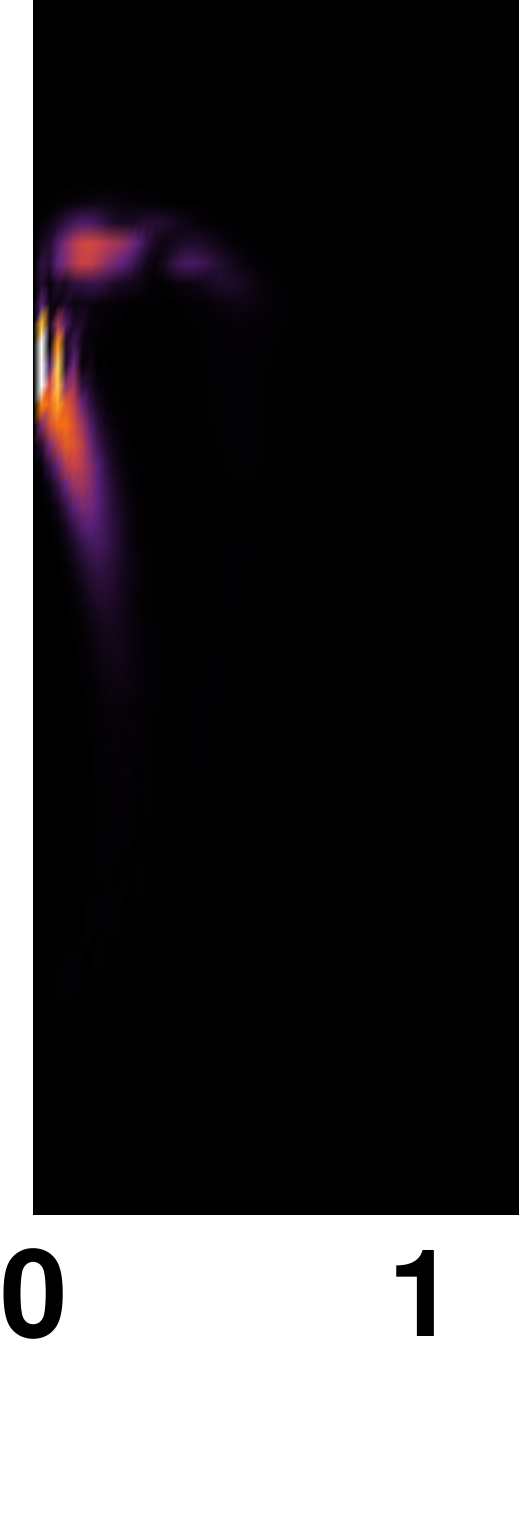}
\caption{(color online) Probability densities in the configuration space for different moments in time, i.e. $\omega t=\pi,\frac{3}{2}\pi,2\pi,\frac{5\pi}{2}$ from left to right, for two Floquet eigenstates. Top panels are related to the eigenstate shown in Fig.~\ref{3Dloc_in_time}(a) while bottom panels to the eigenstate presented in Fig.~\ref{3Dloc_in_time}(b). Panels show cuts along an arbitrary plane containing the $z$ axis multiplied by $\rho$ to simulate the density in cylindrical coordinates. The $\rho$ on the horizontal axis is either $x$ or $y$ or any other direction in the $xy$ plane. The axis scaling factors are identical. Note that in bottom panels a localized electron evolves along an elliptical Kepler orbit whose major axis is not precisely oriented along the $z$ axis. Consequently the axial symmetry of the system results in two parts of the orbits visible in the plots.}
\label{den_config}
\end{figure}

Let us begin with the classical description and perform the canonical transformation to the action angle variables $(J,\theta,L,\Psi)$. The canonically conjugate variables $J$ and $\theta$ are the same as in the 1D model, i.e. they describe the energy of an unperturbed electron and its position on a Kepler ellipse, respectively. The total angular momentum $L$ is conjugate to $\Psi$ and the latter is the angle between the major axis of a Kepler ellipse and the $z$ axis \cite{sacha98}. In the frame moving with an electron, i.e. $\Theta=\theta-\omega t$ and $P=J-J_0$ where $J_0=\omega^{-1/3}$, $\Theta$ and $P$ vary slowly in the vicinity of a resonant orbit, i.e. when $P\approx 0$. The total angular momentum $L$ and the angle $\Psi$ are also slowly varying variables if $FJ_0^4\ll 1$ and $F_sJ_0^4\ll 1$ \cite{delande02}. Thus, similarly as in the 1D model, we may apply the secular approximation and obtain the classical effective Hamiltonian in the rotating frame by averaging the original Hamiltonian over time that yields
\begin{align}
H_{\rm eff}&=&\frac{P^2}{2m_{\rm eff}}+FJ_0^2\sum_{k\ne 0}U_k(L,\Psi)f_{-k}e^{ik\Theta} \cr
&&+F_sJ_0^2\frac{3\tilde e}{2}\cos\Psi, 
\label{heff3d}
\end{align}
where
\bea
U_k(L,\Psi)=\frac{{\cal J}_k'(k\tilde e)}{k}\cos\Psi+i\frac{\sqrt{1-\tilde e^2}}{k\tilde e}{\cal J}_k(k\tilde e)\sin\Psi, \cr 
\label{uk}
\eea
and $\tilde e=\sqrt{1-L^2/J_0^2}$ is the eccentricity of a Kepler ellipse. Similarly as in (\ref{heff}) the constant term $-\frac{3}{2J_0^2}$ has been omitted.

The Hamiltonian (\ref{heff3d}) describes electronic motion on a resonant Kepler orbit and a slow precession of the orbit itself, i.e. slow changes of the variables $L$ and $\Psi$. The motion in the $(P,\Theta)$ space effectively decouples from the slow motion in the $(L,\Psi)$ space and the semiclassical quantization of the system can be performed in the spirit of the Born-Oppenheimer approximation \cite{sacha98,delande02}. That is, first quantize the $P$ and $\Theta$ variables, for frozen $L$ and $\Psi$, and then use the results to construct an effective Hamiltonian for $L$ and $\Psi$ which are subsequently qunatized. This procedure is particularly simple if we are interested in eigenstates which are concentrated, in the semiclassical picture, around a stable fixed point $(L_0,\Psi_0)$ in the $(L,\Psi)$ phase space. Then, we may treat (\ref{heff3d}) as a 1D Hamiltonian where $L=L_0$ and $\Psi=\Psi_0$ are fixed parameters. 

In Sec.~\ref{s1d} we have analyzed Anderson localization of an electron on an orbit degenerated into a line within the 1D model. However, such an orbit is not stable in the 3D description if only a linearly polarized microwave field is applied. In order to change the structure of the classical $(L,\Psi)$  phase space and make $(L_0=0,\Psi_0=0)$ a stable fixed point we have included the static electric field in (\ref{h3d}). If $F_s$ is sufficiently big the orbit with $L_0=0$ oriented along the microwave field polarization axis does not precess and it keeps its shape in time \cite{sacha98,delande02}, see Fig.~\ref{lpsi}. Then, the coefficients (\ref{uk}) become
\be
U_k(0,0)=\frac{{\cal J}_k'(k)}{k},
\ee
and the Hamiltonian (\ref{heff3d}) reduces to the 1D effective Hamiltonian (\ref{heff}). With $f_{-k}$ like in (\ref{fkcond}) we reproduce the previous 1D predictions. This is illustrated in Fig.~\ref{sos} where we present the phase space portrait generated by the 1D secular Hamiltonian (\ref{heffs}) and the stroboscopic picture of the $(P,\Theta)$ phase space obtained in the full 3D classical evolution. 

Quantum results that we will present in the following are achieved by means of the quantum version of the 3D secular Hamiltonian,
\bea
\langle n',l'|\hat H_{\rm eff}|n,l\rangle&=&\left(-\frac{1}{2n^2}-n\omega\right)\delta_{nn'}\delta_{ll'}
\cr
&&+\langle n',l'|z|n,l\rangle \left(Ff_{n-n'}-F_s\delta_{nn'}\right),
\cr &&
\label{qheff_3d}
\eea
where $|n,l\rangle$ is a hydrogenic eigenstate with the principal quantum number $n$, total angular momentum $l$ and the projection of the angular momentum on the $z$ axis equal zero, cf. the 1D counterpart (\ref{qheff}). Diagonalization of (\ref{qheff_3d}) results in a bunch of quasi-energy levels as presented in Fig.~\ref{leveldyn} versus $F_s$. In order to identify desired eigenstates it is very helpful to perform semiclassical quantization of the Hamiltonian (\ref{heff3d}) within the Born-Oppenheimer approximation --- the obtained quasi-energy levels follow very closely desire levels of the Hamiltonian (\ref{qheff_3d}), see Fig.~\ref{leveldyn}. 

In Fig.~\ref{3Dloc_in_time} we present time evolution of the probability density for the detection of an electron at the position of a nucleus of an H atom for two different Floquet eigenstates related to two different values of the static electric field amplitude $F_s$. Panel (a) corresponds to the parameters for which an electron Anderson localizes along an orbit degenerated into a line which is the 3D counterpart of the case described in Sec.~\ref{s1d} in the 1D model, cf. Fig.~\ref{prob1Da}. Panel (b) is related to the case where the classical fixed point is located at $(L/n_0\approx 0.4,\Psi=0.06\pi)$, see Fig.~\ref{lpsi}. Then, an electron localizes along an elliptical trajectory. Evolution of the probability densities in the configuration space is illustrated in Fig.~\ref{den_config} where one can see different shapes of the orbits a localized electron moves on. The major axis of an elliptical orbit visible in bottom panels of Fig.~\ref{den_config} is not exactly oriented along the $z$ axis. Therefore, due to the axial symmetry, one can see two parts of an ellipse, i.e. a part corresponding to, e.g., $x>0$ and a part related to $x<0$ but reflected with respect to $x=0$ plane.

\section{Conclusions}
\label{conl}

Anderson localization in the time domain is a localization phenomenon that can be observed in time evolution of a system exposed to a fluctuating force. Probability density for detection of a system at a fixed position in the configuration space becomes exponentially localized around a certain moment of time due to the presence of {\it disorder} in time \cite{Sacha2015a,sacha_delande16,DelandeMoralesSacha}.

In the present publication we show that a Rydberg atom perturbed by a fluctuating microwave field constitutes a suitable system for realization of Anderson localization in the time domain. Driving of a Rydberg electron by a superposition of a monochromatic microwave field and its few harmonics with random relative phases leads to Anderson localization of an electron along a classical Kepler orbit. That is, the probability for measurement of the electron, e.g., close to the nucleus is exponentially localized around a certain moment of time. 

We analyze a Rydberg atom and present a range of parameters for which Anderson localization in the time domain can be observed experimentally. It seems that the phenomenon can be realized in a laboratory very soon. There are two groups which succeeded in creation of the so-called non-spreading wavepackets in Rydberg atoms driven by microwave fields \cite{maeda04,maeda07,maeda09,wyker12}. Realization of the Anderson localization in time requires modification of the microwave fields only, i.e., switching from the monochromatic fields to fields that fluctuate in time. It would be the first experimental demonstration of Anderson localization in the time domain.


\section*{Acknowledgments}

Support of the National Science Centre, Poland via project No.2016/20/W/ST4/00314
is acknowledged.

%


\begin{thebibliography}{30}%
\makeatletter
\providecommand \@ifxundefined [1]{%
 \@ifx{#1\undefined}
}%
\providecommand \@ifnum [1]{%
 \ifnum #1\expandafter \@firstoftwo
 \else \expandafter \@secondoftwo
 \fi
}%
\providecommand \@ifx [1]{%
 \ifx #1\expandafter \@firstoftwo
 \else \expandafter \@secondoftwo
 \fi
}%
\providecommand \natexlab [1]{#1}%
\providecommand \enquote  [1]{``#1''}%
\providecommand \bibnamefont  [1]{#1}%
\providecommand \bibfnamefont [1]{#1}%
\providecommand \citenamefont [1]{#1}%
\providecommand \href@noop [0]{\@secondoftwo}%
\providecommand \href [0]{\begingroup \@sanitize@url \@href}%
\providecommand \@href[1]{\@@startlink{#1}\@@href}%
\providecommand \@@href[1]{\endgroup#1\@@endlink}%
\providecommand \@sanitize@url [0]{\catcode `\\12\catcode `\$12\catcode
  `\&12\catcode `\#12\catcode `\^12\catcode `\_12\catcode `\%12\relax}%
\providecommand \@@startlink[1]{}%
\providecommand \@@endlink[0]{}%
\providecommand \url  [0]{\begingroup\@sanitize@url \@url }%
\providecommand \@url [1]{\endgroup\@href {#1}{\urlprefix }}%
\providecommand \urlprefix  [0]{URL }%
\providecommand \Eprint [0]{\href }%
\providecommand \doibase [0]{http://dx.doi.org/}%
\providecommand \selectlanguage [0]{\@gobble}%
\providecommand \bibinfo  [0]{\@secondoftwo}%
\providecommand \bibfield  [0]{\@secondoftwo}%
\providecommand \translation [1]{[#1]}%
\providecommand \BibitemOpen [0]{}%
\providecommand \bibitemStop [0]{}%
\providecommand \bibitemNoStop [0]{.\EOS\space}%
\providecommand \EOS [0]{\spacefactor3000\relax}%
\providecommand \BibitemShut  [1]{\csname bibitem#1\endcsname}%
\let\auto@bib@innerbib\@empty
\bibitem [{\citenamefont {Anderson}(1958)}]{Anderson1958}%
  \BibitemOpen
  \bibfield  {author} {\bibinfo {author} {\bibfnamefont {P.~W.}\ \bibnamefont
  {Anderson}},\ }\href {\doibase 10.1103/PhysRev.109.1492} {\bibfield
  {journal} {\bibinfo  {journal} {Phys. Rev.}\ }\textbf {\bibinfo {volume}
  {109}},\ \bibinfo {pages} {1492} (\bibinfo {year} {1958})}\BibitemShut
  {NoStop}%
\bibitem{tiggelen99}
van Tiggelen, B.A. in {\it Diffuse Waves in Complex Media},
edited by J.-P. Fouque, NATO Advanced Study Institutes,
Ser. C, Vol. 531 (Kluwer, Dordrecht, 1999).

\bibitem{Lagendijk2009} Lagendijk, A., van Tiggelen, B.A. and Wiersma, D.S. Fifty years of Anderson localization. {\it Phys. Today} \textbf{62}, 24 (2009).
\bibitem [{\citenamefont {M\"uller}\ and\ \citenamefont
  {Delande}(2011)}]{MuellerDelande:Houches:2009}%
  \BibitemOpen
  \bibfield  {author} {\bibinfo {author} {\bibfnamefont {C.~A.}\ \bibnamefont
  {M\"uller}}\ and\ \bibinfo {author} {\bibfnamefont {D.}~\bibnamefont
  {Delande}},\ }\enquote {\bibinfo {title} {{Disorder and interference:
  localization phenomena}},}\ in\ \href {\doibase
  DOI:10.1093/acprof:oso/9780199603657.003.0009} {\emph {\bibinfo {booktitle}
  {Lecture Notes of the Les Houches Summer School in Singapore: Ultracold Gases
  and Quantum Information}}},\ Vol.~\bibinfo {volume} {91}\ (\bibinfo
  {publisher} {Oxford Scholarship},\ \bibinfo {year} {2011})\ Chap.~\bibinfo
  {chapter} {9},\ \Eprint {http://arxiv.org/abs/1005.0915} {arXiv:1005.0915}
  \BibitemShut {NoStop}%
\bibitem [{\citenamefont {Moore}\ \emph {et~al.}(1995)\citenamefont {Moore},
  \citenamefont {Robinson}, \citenamefont {Bharucha}, \citenamefont
  {Sundaram},\ and\ \citenamefont
  {Raizen}}]{Moore:AtomOpticsRealizationQKR:PRL95}%
  \BibitemOpen
  \bibfield  {author} {\bibinfo {author} {\bibfnamefont {F.~L.}\ \bibnamefont
  {Moore}}, \bibinfo {author} {\bibfnamefont {J.~C.}\ \bibnamefont {Robinson}},
  \bibinfo {author} {\bibfnamefont {C.~F.}\ \bibnamefont {Bharucha}}, \bibinfo
  {author} {\bibfnamefont {B.}~\bibnamefont {Sundaram}}, \ and\ \bibinfo
  {author} {\bibfnamefont {M.~G.}\ \bibnamefont {Raizen}},\ }\href {\doibase
  10.1103/PhysRevLett.75.4598} {\bibfield  {journal} {\bibinfo  {journal}
  {Phys. Rev. Lett.}\ }\textbf {\bibinfo {volume} {75}},\ \bibinfo {pages}
  {4598} (\bibinfo {year} {1995})}\BibitemShut {NoStop}%
\bibitem [{\citenamefont {Fishman}\ \emph {et~al.}(1982)\citenamefont
  {Fishman}, \citenamefont {Grempel},\ and\ \citenamefont
  {Prange}}]{Fishman:LocDynAnders:PRL82}%
  \BibitemOpen
  \bibfield  {author} {\bibinfo {author} {\bibfnamefont {S.}~\bibnamefont
  {Fishman}}, \bibinfo {author} {\bibfnamefont {D.~R.}\ \bibnamefont
  {Grempel}}, \ and\ \bibinfo {author} {\bibfnamefont {R.~E.}\ \bibnamefont
  {Prange}},\ }\href {\doibase 10.1103/PhysRevLett.49.509} {\bibfield
  {journal} {\bibinfo  {journal} {Phys. Rev. Lett.}\ }\textbf {\bibinfo
  {volume} {49}},\ \bibinfo {pages} {509} (\bibinfo {year} {1982})}\BibitemShut
  {NoStop}%
\bibitem [{\citenamefont {Casati}\ \emph {et~al.}(1989)\citenamefont {Casati},
  \citenamefont {Guarneri},\ and\ \citenamefont
  {Shepelyansky}}]{Casati:IncommFreqsQKR:PRL89}%
  \BibitemOpen
  \bibfield  {author} {\bibinfo {author} {\bibfnamefont {G.}~\bibnamefont
  {Casati}}, \bibinfo {author} {\bibfnamefont {I.}~\bibnamefont {Guarneri}}, \
  and\ \bibinfo {author} {\bibfnamefont {D.~L.}\ \bibnamefont {Shepelyansky}},\
  }\href {\doibase 10.1103/PhysRevLett.62.345} {\bibfield  {journal} {\bibinfo
  {journal} {Phys. Rev. Lett.}\ }\textbf {\bibinfo {volume} {62}},\ \bibinfo
  {pages} {345} (\bibinfo {year} {1989})}\BibitemShut {NoStop}%
\bibitem [{\citenamefont {Lemari\'e}\ \emph {et~al.}(2009)\citenamefont
  {Lemari\'e}, \citenamefont {Chab\'e}, \citenamefont {Szriftgiser},
  \citenamefont {Garreau}, \citenamefont {Gr\'emaud},\ and\ \citenamefont
  {Delande}}]{Lemarie:Anderson3D:PRA09}%
  \BibitemOpen
  \bibfield  {author} {\bibinfo {author} {\bibfnamefont {G.}~\bibnamefont
  {Lemari\'e}}, \bibinfo {author} {\bibfnamefont {J.}~\bibnamefont {Chab\'e}},
  \bibinfo {author} {\bibfnamefont {P.}~\bibnamefont {Szriftgiser}}, \bibinfo
  {author} {\bibfnamefont {J.~C.}\ \bibnamefont {Garreau}}, \bibinfo {author}
  {\bibfnamefont {B.}~\bibnamefont {Gr\'emaud}}, \ and\ \bibinfo {author}
  {\bibfnamefont {D.}~\bibnamefont {Delande}},\ }\href {\doibase
  10.1103/PhysRevA.80.043626} {\bibfield  {journal} {\bibinfo  {journal} {Phys.
  Rev. A}\ }\textbf {\bibinfo {volume} {80}},\ \bibinfo {pages} {043626}
  (\bibinfo {year} {2009})},\ \Eprint {http://arxiv.org/abs/0907.3411}
  {arXiv:0907.3411 [quant-ph]} \BibitemShut {NoStop}%
\bibitem [{\citenamefont {Manai}\ \emph {et~al.}(2015)\citenamefont {Manai},
  \citenamefont {Cl{\'e}ment}, \citenamefont {Chicireanu}, \citenamefont
  {Hainaut}, \citenamefont {Garreau}, \citenamefont {Szriftgiser},\ and\
  \citenamefont {Delande}}]{Manai:Anderson2DKR:PRL15}%
  \BibitemOpen
  \bibfield  {author} {\bibinfo {author} {\bibfnamefont {I.}~\bibnamefont
  {Manai}}, \bibinfo {author} {\bibfnamefont {J.-F.}\ \bibnamefont
  {Cl{\'e}ment}}, \bibinfo {author} {\bibfnamefont {R.}~\bibnamefont
  {Chicireanu}}, \bibinfo {author} {\bibfnamefont {C.}~\bibnamefont {Hainaut}},
  \bibinfo {author} {\bibfnamefont {J.~C.}\ \bibnamefont {Garreau}}, \bibinfo
  {author} {\bibfnamefont {P.}~\bibnamefont {Szriftgiser}}, \ and\ \bibinfo
  {author} {\bibfnamefont {D.}~\bibnamefont {Delande}},\ }\href {\doibase
  10.1103/PhysRevLett.115.240603} {\bibfield  {journal} {\bibinfo  {journal}
  {Phys. Rev. Lett.}\ }\textbf {\bibinfo {volume} {115}},\ \bibinfo {pages}
  {240603} (\bibinfo {year} {2015})}\BibitemShut {NoStop}%
\bibitem [{\citenamefont {Sacha}(2015{\natexlab{b}})}]{Sacha2015a}%
  \BibitemOpen
  \bibfield  {author} {\bibinfo {author} {\bibfnamefont {K.}~\bibnamefont
  {Sacha}},\ }\href@noop {} {\bibfield  {journal} {\bibinfo  {journal} {Sci.
  Rep.}\ }\textbf {\bibinfo {volume} {5}},\ \bibinfo {pages} {10787} (\bibinfo
  {year} {2015}{\natexlab{b}})}\BibitemShut {NoStop}%
\bibitem{sacha_delande16}  
  K. Sacha and D. Delande, Phys. Rev. A {\bf 94}, 023633 (2016).
\bibitem{DelandeMoralesSacha}
  D. Delande, L. Morales-Molina, and K. Sacha, arxiv:1702.03591.
\bibitem [{\citenamefont {Wilczek}(2012)}]{Wilczek2012}%
  \BibitemOpen
  \bibfield  {author} {\bibinfo {author} {\bibfnamefont {F.}~\bibnamefont
  {Wilczek}},\ }\href {\doibase 10.1103/PhysRevLett.109.160401} {\bibfield
  {journal} {\bibinfo  {journal} {Phys. Rev. Lett.}\ }\textbf {\bibinfo
  {volume} {109}},\ \bibinfo {pages} {160401} (\bibinfo {year}
  {2012})}\BibitemShut {NoStop}%
\bibitem [{\citenamefont {Li}\ \emph {et~al.}(2012{\natexlab{a}})\citenamefont
  {Li}, \citenamefont {Gong}, \citenamefont {Yin}, \citenamefont {Quan},
  \citenamefont {Yin}, \citenamefont {Zhang}, \citenamefont {Duan},\ and\
  \citenamefont {Zhang}}]{Li2012}%
  \BibitemOpen
  \bibfield  {author} {\bibinfo {author} {\bibfnamefont {T.}~\bibnamefont
  {Li}}, \bibinfo {author} {\bibfnamefont {Z.-X.}\ \bibnamefont {Gong}},
  \bibinfo {author} {\bibfnamefont {Z.-Q.}\ \bibnamefont {Yin}}, \bibinfo
  {author} {\bibfnamefont {H.~T.}\ \bibnamefont {Quan}}, \bibinfo {author}
  {\bibfnamefont {X.}~\bibnamefont {Yin}}, \bibinfo {author} {\bibfnamefont
  {P.}~\bibnamefont {Zhang}}, \bibinfo {author} {\bibfnamefont {L.-M.}\
  \bibnamefont {Duan}}, \ and\ \bibinfo {author} {\bibfnamefont
  {X.}~\bibnamefont {Zhang}},\ }\href {\doibase 10.1103/PhysRevLett.109.163001}
  {\bibfield  {journal} {\bibinfo  {journal} {Phys. Rev. Lett.}\ }\textbf
  {\bibinfo {volume} {109}},\ \bibinfo {pages} {163001} (\bibinfo {year}
  {2012}{\natexlab{a}})}\BibitemShut {NoStop}%
\bibitem [{\citenamefont {Chernodub}(2013)}]{Chernodub2013}%
  \BibitemOpen
  \bibfield  {author} {\bibinfo {author} {\bibfnamefont {M.~N.}\ \bibnamefont
  {Chernodub}},\ }\href {\doibase 10.1103/PhysRevD.87.025021} {\bibfield
  {journal} {\bibinfo  {journal} {Phys. Rev. D}\ }\textbf {\bibinfo {volume}
  {87}},\ \bibinfo {pages} {025021} (\bibinfo {year} {2013})}\BibitemShut
  {NoStop}%
\bibitem [{\citenamefont {Wilczek}(2013{\natexlab{a}})}]{Wilczek2013}%
  \BibitemOpen
  \bibfield  {author} {\bibinfo {author} {\bibfnamefont {F.}~\bibnamefont
  {Wilczek}},\ }\href {\doibase 10.1103/PhysRevLett.111.250402} {\bibfield
  {journal} {\bibinfo  {journal} {Phys. Rev. Lett.}\ }\textbf {\bibinfo
  {volume} {111}},\ \bibinfo {pages} {250402} (\bibinfo {year}
  {2013}{\natexlab{a}})}\BibitemShut {NoStop}%
\bibitem [{\citenamefont {Bruno}(2013{\natexlab{a}})}]{Bruno2013}%
  \BibitemOpen
  \bibfield  {author} {\bibinfo {author} {\bibfnamefont {P.}~\bibnamefont
  {Bruno}},\ }\href {\doibase 10.1103/PhysRevLett.110.118901} {\bibfield
  {journal} {\bibinfo  {journal} {Phys. Rev. Lett.}\ }\textbf {\bibinfo
  {volume} {110}},\ \bibinfo {pages} {118901} (\bibinfo {year}
  {2013}{\natexlab{a}})}\BibitemShut {NoStop}%
\bibitem [{\citenamefont {Wilczek}(2013{\natexlab{b}})}]{Wilczek2013a}%
  \BibitemOpen
  \bibfield  {author} {\bibinfo {author} {\bibfnamefont {F.}~\bibnamefont
  {Wilczek}},\ }\href {\doibase 10.1103/PhysRevLett.110.118902} {\bibfield
  {journal} {\bibinfo  {journal} {Phys. Rev. Lett.}\ }\textbf {\bibinfo
  {volume} {110}},\ \bibinfo {pages} {118902} (\bibinfo {year}
  {2013}{\natexlab{b}})}\BibitemShut {NoStop}%
\bibitem [{\citenamefont {Bruno}(2013{\natexlab{b}})}]{Bruno2013a}%
  \BibitemOpen
  \bibfield  {author} {\bibinfo {author} {\bibfnamefont {P.}~\bibnamefont
  {Bruno}},\ }\href {\doibase 10.1103/PhysRevLett.111.029301} {\bibfield
  {journal} {\bibinfo  {journal} {Phys. Rev. Lett.}\ }\textbf {\bibinfo
  {volume} {111}},\ \bibinfo {pages} {029301} (\bibinfo {year}
  {2013}{\natexlab{b}})}\BibitemShut {NoStop}%
\bibitem [{\citenamefont {Li}\ \emph {et~al.}(2012{\natexlab{b}})\citenamefont
  {Li}, \citenamefont {Gong}, \citenamefont {Yin}, \citenamefont {Quan},
  \citenamefont {Yin}, \citenamefont {Zhang}, \citenamefont {Duan},\ and\
  \citenamefont {Zhang}}]{Li2012a}%
  \BibitemOpen
  \bibfield  {author} {\bibinfo {author} {\bibfnamefont {T.}~\bibnamefont
  {Li}}, \bibinfo {author} {\bibfnamefont {Z.-X.}\ \bibnamefont {Gong}},
  \bibinfo {author} {\bibfnamefont {Z.-Q.}\ \bibnamefont {Yin}}, \bibinfo
  {author} {\bibfnamefont {H.~T.}\ \bibnamefont {Quan}}, \bibinfo {author}
  {\bibfnamefont {X.}~\bibnamefont {Yin}}, \bibinfo {author} {\bibfnamefont
  {P.}~\bibnamefont {Zhang}}, \bibinfo {author} {\bibfnamefont {L.-M.}\
  \bibnamefont {Duan}}, \ and\ \bibinfo {author} {\bibfnamefont
  {X.}~\bibnamefont {Zhang}},\ }\href@noop {} {\  (\bibinfo {year}
  {2012}{\natexlab{b}})},\ \Eprint {http://arxiv.org/abs/1212.6959}
  {arXiv:1212.6959} \BibitemShut {NoStop}%
\bibitem [{\citenamefont {Bruno}(2013{\natexlab{c}})}]{Bruno2013b}%
  \BibitemOpen
  \bibfield  {author} {\bibinfo {author} {\bibfnamefont {P.}~\bibnamefont
  {Bruno}},\ }\href {\doibase 10.1103/PhysRevLett.111.070402} {\bibfield
  {journal} {\bibinfo  {journal} {Phys. Rev. Lett.}\ }\textbf {\bibinfo
  {volume} {111}},\ \bibinfo {pages} {070402} (\bibinfo {year}
  {2013}{\natexlab{c}})}\BibitemShut {NoStop}%
\bibitem{Guo2013}
  L. Guo, M. Marthaler, and G. Sch\"on, Phys. Rev. Lett. {\bf 111}, 205303 (2013).
\bibitem [{\citenamefont {Watanabe}\ and\ \citenamefont
  {Oshikawa}(2015)}]{Watanabe2015}%
  \BibitemOpen
  \bibfield  {author} {\bibinfo {author} {\bibfnamefont {H.}~\bibnamefont
  {Watanabe}}\ and\ \bibinfo {author} {\bibfnamefont {M.}~\bibnamefont
  {Oshikawa}},\ }\href {\doibase 10.1103/PhysRevLett.114.251603} {\bibfield
  {journal} {\bibinfo  {journal} {Phys. Rev. Lett.}\ }\textbf {\bibinfo
  {volume} {114}},\ \bibinfo {pages} {251603} (\bibinfo {year}
  {2015})}\BibitemShut {NoStop}%
\bibitem [{\citenamefont {Sacha}(2015{\natexlab{a}})}]{Sacha2015}%
  \BibitemOpen
  \bibfield  {author} {\bibinfo {author} {\bibfnamefont {K.}~\bibnamefont
  {Sacha}},\ }\href {\doibase 10.1103/PhysRevA.91.033617} {\bibfield  {journal}
  {\bibinfo  {journal} {Phys. Rev. A}\ }\textbf {\bibinfo {volume} {91}},\
  \bibinfo {pages} {033617} (\bibinfo {year} {2015}{\natexlab{a}})}\BibitemShut
  {NoStop}%
\bibitem{khemani16}
V. Khemani, A. Lazarides, R. Moessner, and S. L. Sondhi, Phys. Rev. Lett. {\bf 116}, 250401 (2016).  
\bibitem{else16}  
 D. V. Else, B. Bauer, and C. Nayak, Phys. Rev. Lett. {\bf 117}, 090402 (2016).
\bibitem{Keyserlingk16}
  C. von Keyserlingk, V. Khemani, and S. Sondhi, Phys. Rev. B {\bf 94}, 085112 (2016).  
\bibitem{yao17}
  N. Y. Yao, A. C. Potter, I. D. Potirniche, and A. Vishwanath, Phys. Rev. Lett. {\bf 118}, 030401 (2017).
\bibitem{Weidinger16}
 S. A. Weidinger, and M. Knap,  arXiv:1609.09089.
\bibitem{zhang16}
 J. Zhang, P. W. Hess, A. Kyprianidis, P. Becker, A. Lee, J. Smith, G. Pagano, I.-D. Potirniche, A. C. Potter, A. Vishwanath, N. Y. Yao, C. Monroe, Nature {\bf 543}, 217 (2017).
\bibitem{choi16}
   S. Choi, J. Choi, R. Landig, G. Kucsko, H. Zhou, J. Isoya, F. Jelezko, S. Onoda, H. Sumiya, V. Khemani, C. von Keyserlingk, N. Y. Yao, E. Demler, M. D. Lukin, Nature {\bf 543}, 221 (2017).
\bibitem{Syrwid17}
 A. Syrwid, J. Zakrzewski, and K. Sacha, arXiv:1702.05006.
\bibitem{zas} G. P. Berman and G. M. Zaslavsky, Phys. Lett. A {\bf 61},
295 (1977).

\bibitem{hen+92} J. Henkel and M. Holthaus, Phys.\ Rev.\ {\bf A45}, 1978 (1992).

\bibitem{hol95} M. Holthaus, Chaos, Solitons and Fractals {\bf 5}, 1143 (1995).

\bibitem{del+94} D. Delande and A. Buchleitner,  Adv.\ At. Mol. Opt. Phys. 
{\bf 35}, 85 (1994).

\bibitem{ibb}          I. Bialynicki-Birula, M. Kali\'nski, and J. H. Eberly,
 Phys.\ Rev.\ Lett.\ {\bf 73}, 1777 (1994).

\bibitem{buch+95} A. Buchleitner and D. Delande,
 Phys.\ Rev.\ Lett.\ {\bf 75}, 1487 (1995).

\bibitem{dzb95} D. Delande, J. Zakrzewski, and A. Buchleitner,
                 Europhys.\ Lett.\ {\bf 32}, 107 (1995).

\bibitem{zdb95} J. Zakrzewski, D. Delande, and A.
 Buchleitner, Phys.\ Rev.\ Lett.\ {\bf 75}, 4015 (1995).
 
\bibitem{sacha98}
 K. Sacha, J. Zakrzewski and D. Delande, Europ. Phys. J. D, {\bf 1}, 231 (1998).
 
\bibitem{delande02}
D. Delande, K. Sacha, and J. Zakrzewski, Acta Physica Polonica B, {\bf 33}, 2097 (2002), arXiv:quant-ph/0206033.
 
\bibitem{maeda04}
H. Maeda and T. F. Gallagher. Phys. Rev. Lett. {\bf 92}, 133004 (2004).

\bibitem{maeda07}
H. Maeda and T. F. Gallagher, Phys. Rev. A {\bf 75}, 033410 (2007).

\bibitem{maeda09}
H. Maeda, J. H. Gurian, and T. F. Gallagher, Phys. Rev. Lett. 102, 103001 (2009).

\bibitem{wyker12}
B. Wyker, S. Ye, F. B. Dunning, S. Yoshida, C. O. Reinhold, and J. Burgd\"orfer,
Phys. Rev. Lett. {\bf 108}, 043001 (2012).

\bibitem [{\citenamefont {Buchleitner}\ \emph {et~al.}(2002)\citenamefont
  {Buchleitner}, \citenamefont {Delande},\ and\ \citenamefont
  {Zakrzewski}}]{Buchleitner2002}%
  \BibitemOpen
  \bibfield  {author} {\bibinfo {author} {\bibfnamefont {A.}~\bibnamefont
  {Buchleitner}}, \bibinfo {author} {\bibfnamefont {D.}~\bibnamefont
  {Delande}}, \ and\ \bibinfo {author} {\bibfnamefont {J.}~\bibnamefont
  {Zakrzewski}},\ }\href
  {http://www.sciencedirect.com/science/article/pii/S0370157302002703}
  {\bibfield  {journal} {\bibinfo  {journal} {Phys. Rep.}\ }\textbf
  {\bibinfo {volume} {368}},\ \bibinfo {pages} {409} (\bibinfo {year}
  {2002})}\BibitemShut {NoStop}%
\bibitem{Lichtenberg_s}  
A. J. Lichtenberg and M. A. Lieberman, {\it Regular and Stochastic Motion}, Applied Mathematical Sciences Vol. 38, edited by F. John {\it et al.} (Springer, Berlin 1983).

\bibitem [{\citenamefont {Kuhn}\ \emph {et~al.}(2007)\citenamefont {Kuhn},
  \citenamefont {Sigwarth}, \citenamefont {Miniatura}, \citenamefont
  {Delande},\ and\ \citenamefont {M\"uller}}]{Kuhn:Speckle:NJP07}%
  \BibitemOpen
  \bibfield  {author} {\bibinfo {author} {\bibfnamefont {R.~C.}\ \bibnamefont
  {Kuhn}}, \bibinfo {author} {\bibfnamefont {O.}~\bibnamefont {Sigwarth}},
  \bibinfo {author} {\bibfnamefont {C.}~\bibnamefont {Miniatura}}, \bibinfo
  {author} {\bibfnamefont {D.}~\bibnamefont {Delande}}, \ and\ \bibinfo
  {author} {\bibfnamefont {C.~A.}\ \bibnamefont {M\"uller}},\ }\href {\doibase
  10.1088/1367-2630/9/6/161} {\bibfield  {journal} {\bibinfo  {journal} {New J.
  Phys.}\ }\textbf {\bibinfo {volume} {9}},\ \bibinfo {pages} {161} (\bibinfo
  {year} {2007})}\BibitemShut {NoStop}%
\bibitem{shirley65}
  J. H. Shirley, Phys. Rev. {\bf 138}, B979 (1965).
\bibitem{lugan}
Pierre Lugan. Ultracold Bose gases in random potentials: collective excitations and localization effects. Atomic Physics [physics.atom-ph]. Ecole Polytechnique X, 2010. English.
<tel-00468888>
  
\end{thebibliography}
\end{document}